\documentclass[12pt,a4paper]{article}

\usepackage[margin=1in]{geometry}
\usepackage{graphicx}
\usepackage{float}
\usepackage{booktabs}
\usepackage{amsmath}
\usepackage{caption}
\usepackage[utf8]{inputenc}
\usepackage{amssymb}
\usepackage{hyperref}
\usepackage{amsmath}
\usepackage{times}
\usepackage{xcolor}
\usepackage{authblk}
\usepackage{tabularx}
\usepackage{booktabs}
\usepackage{subcaption}

\hypersetup{
    colorlinks=true,
    linkcolor=blue,
    urlcolor=blue,
    citecolor=blue
}

\providecommand{\keywords}[1]
{
  \small	
  \textbf{\textit{Keywords---}} #1
}

\title{\textbf{Neural Networks as Physics-Consistent Surrogates: An \textit{Explainable AI} Validation Framework for Learning Constitutive Relations }}
\author[1]{Chandana Pati}
\author[1,$\dagger$]{S. M. Mallikarjunaiah}
\affil[1]{Department of Mathematics \& Statistics, Texas A\&M University-Corpus Christi, Texas- 78412, USA}
\affil[$\dagger$]{Corresponding author}
\affil[ ]{\textit{E-mail addresses:} \texttt{cpati@islander.tamucc.edu} (Chandana Pati), \texttt{M.Muddamallappa@tamucc.edu} (S.M. Mallikarjunaiah)}
\date{}

\usepackage{algorithm}
\usepackage{algpseudocode}
\usepackage[utf8]{inputenc}

\begin{document}
\maketitle

\begin{abstract}
This paper presents a Physics-\textit{Explainable AI}  (XAI) framework to validate and interpret neural networks for the constitutive modeling of solid materials. The study bridges the gap between data-driven models and continuum mechanics by applying a suite of explainability methods to neural networks trained on three distinct material behaviors: hyperelasticity (\textit{Mooney-Rivlin}), elastoplasticity (\textit{Chaboche}), and viscoelasticity (\textit{Fractional Zener}). First, high-fidelity surrogate models, including dense feed-forward networks, LSTMs, and GRUs, are trained on synthetically generated data to accurately capture complex material responses. The core of the work then employs XAI techniques to "open the black box" and confirm that the networks learn physically meaningful principles. For hyperelasticity, gradient-based attributions (\textit{Grad Input} (GI)) successfully match the analytical tangent modulus, proving the network learned material stiffness. For elastoplasticity, \textit{SHapley Additive exPlanations} (SHAP) and \textit{Principal Component Analysis} (PCA)  demonstrate the \textit{Recurrent Neural Network} (RNN) internalizes path-dependent memory, with SHAP identifying \textit{plastic strain} as the dominant feature governing the stress prediction. For viscoelasticity, latent-space and wavelet analyses of the \textit{Gated Recurrent Unit. } GRU layers reveal a clear temporal hierarchy, with different layers encoding instantaneous elastic response, intermediate relaxation, and long-term fractional memory. Ultimately, the study demonstrates that the XAI framework can verify that the neural networks are not merely curve-fitting but are, in fact, learning the underlying physical mechanisms of stiffness, history-dependence, and temporal damping.
\end{abstract}

\noindent \keywords{Explainable AI; Constitutive Modeling; Neural Networks; Material Modeling; Hyperelasticity; Elastoplasticity; Viscoelasticity}

\section{Introduction}

The integration of machine learning (ML), particularly deep learning, is poised to revolutionize the field of computational solid mechanics \cite{LeCun2015}. For decades, the development of robust constitutive models has been a cornerstone of mechanical simulation. However, classical phenomenological models, while foundational, often require extensive and costly experimental calibration, struggle to capture extreme nonlinearities, and may fail entirely when faced with complex, non-proportional loading paths or anisotropic material evolution. Artificial neural networks (ANNs), as powerful universal function approximators \cite{Hornik1989}, offer a compelling alternative.

In mechanics, this advantage is being actively explored in two primary domains. First, ANNs serve as highly efficient surrogate models for accelerating expensive, high-fidelity simulations. This includes applications in model order reduction (MOR) \cite{Koeppe2020a} and uncertainty quantification \cite{Bessa2017}, where thousands of simulation calls are rendered computationally feasible. Second, and perhaps more profoundly, ANNs are being used as data-driven constitutive models that can be trained to capture extraordinarily complex material behaviors directly from data \cite{Ghaboussi1991}. Specialized architectures, such as Recurrent Neural Networks (RNNs) and their gated variants (LSTMs and GRUs), are particularly adept at modeling path-dependent and history-dependent phenomena. This has shown great promise for materials like elastoplastic metals \cite{Oeser2009, Heider2020} and viscoelastic polymers \cite{Graf2012}. This data-driven approach promises to bypass the need for deriving complex, analytical material laws and can even bridge scales, learning continuum-level free energy functions directly from atomistic simulations \cite{Teichert2019} or modeling complex phenomena like brittle fracture \cite{Koeppe2017}.

This drive to capture extreme material behavior is part of a larger, long-standing challenge in computational mechanics, and it highlights a fascinating fork in the road for the research community. For decades, the primary path forward has been to build ever-more sophisticated classical models and solvers. This includes deep, foundational work on the very nature of materials, such as implicit constitutive theories \cite{rajagopal2003implicit}, the complex mechanics of mixtures \cite{rajagopal1995mechanics}, and advanced nonlinear elasticity theories \cite{rajagopal2007elasticity}. To actually \textit{solve} these, especially for notoriously difficult problems like fracture, the community has developed a powerful arsenal of specialized numerical techniques. We see this in the development of adaptive phase-field models \cite{yoon2021quasi}, $hp$-adaptive discontinuous Galerkin methods \cite{manohar2024hp}, and intricate $hp$-adaptive finite element (FEM) discretizations \cite{mallikarjunaiah2025hp}. All this effort is aimed at a singular goal: accurately simulating how cracks propagate in novel nonlinear materials, such as those described by strain-limiting elastic theories \cite{manohar2025convergent, mallikarjunaiah2025crack, ghosh2025computational}. But a parallel, more recent approach is emerging, one that uses deep learning not just as a material model, but as a \textit{direct solver} for the governing differential equations. Instead of meticulously crafting FEM code, researchers are now using feedforward neural networks to approximate solutions for complex nonlinear ODEs \cite{martinez2024approximation,venkatachalapathy2023feedforward,venkatachalapathy2023deep} and even singularly perturbed delay differential equations \cite{mallikarjunaiah2023deep}. Both of these streams—one based on refining classical methods, the other on a new computational paradigm—are ultimately trying to get to the same place: a reliable, accurate model of nonlinear physics.

Despite this enormous potential, the primary barrier to the widespread adoption of ANNs in engineering and materials science remains their inherent "black box" nature \cite{Breiman2001}. The internal parameters of a deep neural network, often numbering in the millions, are not readily interpretable in the language of physics or engineering. This opacity leads to a critical lack of trust, which is a non-starter for safety-critical applications in aerospace, biomedical engineering, and civil infrastructure. Without a clear understanding of a model's internal reasoning, it is impossible to perform the rigorous Verification and Validation (V\&V) that these fields mandate \cite{Antolotti2023}. We cannot be certain whether the network has learned the correct, generalizable underlying physical laws or has simply become a "Clever Hans" predictor, exploiting spurious correlations or dataset biases to achieve a correct answer for the wrong reasons \cite{Lapuschkin2019}.

To bridge this "trust gap," two complementary fields have emerged. The first is  \textit{Physics-Informed Neural Networks} (PINNs) \cite{Raissi2019}, which embed \textit{a priori} physical knowledge directly into the learning process. This is most often achieved by incorporating the residual of the governing partial differential equations (PDEs) as a penalty term in the network's loss function. In this way, the network is trained not only to fit the data but also to obey fundamental physical laws, such as the conservation of momentum or energy. This approach, pioneered by \cite{Lagaris1998}, has been shown to dramatically improve data efficiency and ensure the resulting model is physically plausible. It has seen widespread success in fluid dynamics, heat transfer, and increasingly in solid mechanics \cite{Haghighat2021, Karniadakis2021}. However, PINNs are not a silver bullet. They can struggle with stiff PDEs, complex geometries, and high-frequency or chaotic solutions. Furthermore, while they enforce physics, they do not fully solve the interpretability problem; the internal reasoning of the hidden layers, and how they combine information to satisfy the enforced laws, can remain a mystery.

This is where the second, and perhaps more transformative, field of XAI provides a solution \cite{Montavon2018}. Instead of only enforcing physics \textit{a priori}, XAI provides a suite of tools to interpret and explain the network's decisions \textit{a posteriori}. This facilitates a paradigm shift from "physics-informed" to "physics-explaining." It allows us to "open the black box" and perform a scientific interrogation of the model. 

XAI methods can be broadly categorized.  \textit{Attribution methods} seek to identify which inputs were most relevant for a given output. Techniques like Layer-Wise Relevance Propagation (LRP) \cite{Bach2015}, SHapley Additive exPlanations (SHAP) \cite{Lundberg2017}, and Integrated Gradients \cite{Sundararajan2017} can produce "heatmaps" that, in a mechanics context, could answer questions like: "To predict the current stress, did the network correctly 'look at' the accumulated plastic strain?" or "Is the network correctly identifying the yield surface?" Other, model-agnostic methods like LIME \cite{Ribeiro2016} fit simpler, interpretable surrogate models (e.g., a linear model) to the ANN's local decision boundary. A third, and highly relevant, category is  \textit{latent-space analysis}. For RNNs trained on history-dependent materials, we can apply techniques like Principal Component Analysis (PCA) to the network's hidden cell states. This allows us to investigate whether the network has independently learned to encode fundamental physical concepts, such as an internal variable that directly correlates with the "ground truth" plastic strain or a relaxation modulus.

The true advantage of a robust XAI framework is twofold. First, it provides the verification and confidence needed to deploy complex, high-performance, data-driven models in real-world engineering. Second, and more profoundly, it transforms the neural network from a simple regression tool into a powerful partner for scientific discovery. By training a network on raw experimental data—perhaps from complex, multiaxial tests where no clean analytical model exists—and then using XAI to interpret its internal logic, we may discover new, more elegant, or more efficient closed-form solutions for material behavior that were previously unknown \cite{Flaschel2021}.

This work is situated in this exciting new frontier. We aim to develop and apply a comprehensive XAI framework specifically tailored to validate and interpret ANNs for the constitutive modeling of solid materials. By applying this framework to networks trained on fundamental material behaviors—such as hyperelasticity, elastoplasticity, and viscoelasticity—we will demonstrate that XAI can move beyond simple validation. The ultimate goal is to show that these techniques can verify that the networks are not merely curve-fitting, but are, in fact, learning the underlying physical mechanisms of stiffness, history-dependence, and dissipation, effectively transforming them from opaque "black boxes" into interpretable "glass boxes."

\section{Materials and Methods}
This study develops an extended Physics--XAI neural framework for constitutive modeling of solid 
materials \cite{koeppe2022explainable,Karniadakis2021,Raissi2019}. 
The proposed approach integrates physics-based analytical equations 
\cite{ogden1997non,holzapfel2000nonlinear,rajagopal2003implicit} 
with data-driven recurrent neural networks (RNNs) 
\cite{Ghaboussi1991,Koeppe2020a,Oeser2009} 
and gradient-based explainability methods 
\cite{sundararajan2017axiomatic,simonyan2013deep,Bach2015,Montavon2018}. 
Three representative material behaviors---hyperelasticity, elastoplasticity, and viscoelasticity---are 
modeled through their respective rheological analogs 
\cite{ogden1997non,holzapfel2000nonlinear,Oeser2009}. 
Each constitutive model is formulated analytically to generate synthetic datasets 
\cite{Lagaris1998,Raissi2019}, followed by independent neural training 
\cite{LeCun2015,Hornik1989} 
and explainable analysis to identify the encoded physical patterns within latent variables 
\cite{Lapuschkin2019,Montavon2018,Bach2015}.

\subsection{Dataset Generation: Independent Implementation}
All datasets used in this study were independently generated using custom Python scripts rather than 
relying on existing frameworks. Each script reproduces the governing constitutive equations of the 
selected material class and ensures a physically consistent stress--strain--time relationship 
\cite{ogden1997non,holzapfel2000nonlinear,rajagopal2003implicit}. 
Datasets were exported as comma-separated values (.csv) files and visually validated prior to neural 
training. The three implemented constitutive models are summarized below.

\paragraph{(1) Hyperelasticity:}
The hyperelastic dataset represents a Mooney--Rivlin solid extended with Physics--Aware Distillation 
(PAD) to improve stability and interpretability 
\cite{koeppe2022explainable,Lagaris1998,Raissi2019}. 
This model describes nonlinear elastic deformation in incompressible solids under uniaxial loading, 
governed by the classical Mooney--Rivlin stress relation \cite{ogden1997non,holzapfel2000nonlinear}:

\begin{equation}
\sigma = 
2\left( C_1 + \frac{C_2}{\lambda} \right)(\lambda - \lambda^{-2}),
\label{eq:mooney_rivlin}
\end{equation}

where $\sigma$ is the Cauchy stress, $\lambda$ the stretch ratio, and $C_1, C_2$ are material parameters.  
Randomized sampling of $C_1, C_2 \in [0.2,1.0]$ introduces realistic variability across loading paths. 
Cyclic stretch variations between $\lambda = 0.8$ and $\lambda = 1.6$ yield smooth reversible 
stress--strain curves. The PAD procedure later uses the Mooney--Rivlin law defined in 
Eq.~\ref{eq:mooney_rivlin} as a reference model to extract physically interpretable stiffness parameters 
from the trained neural surrogate.

\paragraph{(2) Elastoplasticity:}
The elastoplastic dataset is generated using the Chaboche model, which combines isotropic and kinematic 
hardening to capture cyclic plasticity 
\cite{Ghaboussi1991,Koeppe2020a,Heider2020}. 
The incremental constitutive relations governing the evolution of stress, plastic strain, and backstress are 
given in Eqs.~\ref{eq:chaboche_stress}--\ref{eq:chaboche_backstress}:

\begin{align}
\sigma &= E\left(\varepsilon - \varepsilon^{p}\right),
\label{eq:chaboche_stress} \\
\dot{\varepsilon}^{p} &= 
\frac{1}{\eta}
\left\langle |\sigma - \alpha| - \sigma_y \right\rangle
\,\mathrm{sign}(\sigma - \alpha),
\label{eq:chaboche_flow} \\
\dot{\alpha} &= C \dot{\varepsilon}^{p},
\label{eq:chaboche_backstress}
\end{align}

where $E$ is Young's modulus, $\sigma_y$ is the yield stress, $\alpha$ is the kinematic backstress, 
$C$ is the hardening modulus, and $\eta$ is the viscous regularization parameter.  
Cyclic strain-controlled loading is applied to reproduce yielding, unloading, Bauschinger effects, and 
ratcheting. The full evolution law in 
Eqs.~\ref{eq:chaboche_stress}--\ref{eq:chaboche_backstress} 
serves as the reference model for generating plastic deformation histories, enabling the neural network 
to learn reversible--irreversible transitions and internal variable memory.
For each loading increment, the $\hat{\sigma}$ was computed using the elastic–plastic split defined in 
Eq.~\ref{eq:chaboche_stress}, while the backstress evolution followed the kinematic hardening law 
specified in Eq.~\ref{eq:chaboche_backstress}. Together with the plastic flow rule in 
Eq.~\ref{eq:chaboche_flow}, these relations govern the evolution of plastic strain, backstress, and 
stress asymmetry during cyclic loading. The parameters $E$, $C$, $\gamma$, and $\sigma_y$ were 
sampled within physically realistic ranges to generate diverse loading histories.The resulting stress–strain dataset naturally captured key elastoplastic phenomena—including closed 
hysteresis loops, residual strain accumulation, and the Bauschinger effect—providing a physically rich 
and mechanically accurate foundation for training data-driven constitutive models 
\cite{Ghaboussi1991,Koeppe2020a,Oeser2009}.

\paragraph{(3) Viscoelasticity:}
The viscoelastic dataset is derived from the Fractional Zener model, which incorporates 
fractional-order derivatives to represent long-term relaxation and creep behavior 
\cite{holzapfel2000nonlinear,Teichert2019}. 
The constitutive equation governing viscoelastic stress evolution is given by Eq.~\ref{eq:fractional_zener}:

\begin{equation}
\sigma + \tau^{\beta} D_t^{\beta}\sigma 
= E_1 \varepsilon + E_2 \tau^{\beta} D_t^{\beta}\varepsilon,
\label{eq:fractional_zener}
\end{equation}

where $E_1$ and $E_2$ are elastic moduli, $\tau$ is the characteristic relaxation time, and 
$D_t^{\beta}$ denotes the Caputo fractional derivative of order $\beta \in (0,1)$.  
Random sampling of $\beta$, $E_1$, $E_2$, and $\tau$ introduces variability in relaxation time scales, 
while cyclic, creep, and recovery loading produces temporal sequences exhibiting both instantaneous 
elastic response and delayed viscous relaxation dictated by Eq.~\ref{eq:fractional_zener}.The fractional derivative order $\alpha = 0.6$ governs the strength of hereditary
behavior, interpolating between the purely elastic limit ($\alpha = 0$) and the classical viscous
Newtonian response ($\alpha = 1$), consistent with standard treatments of fractional
viscoelasticity~\cite{holzapfel2000nonlinear,mainardi2022fractional}.

\begin{figure}[H]
    \centering
    \includegraphics[width=0.65\linewidth]{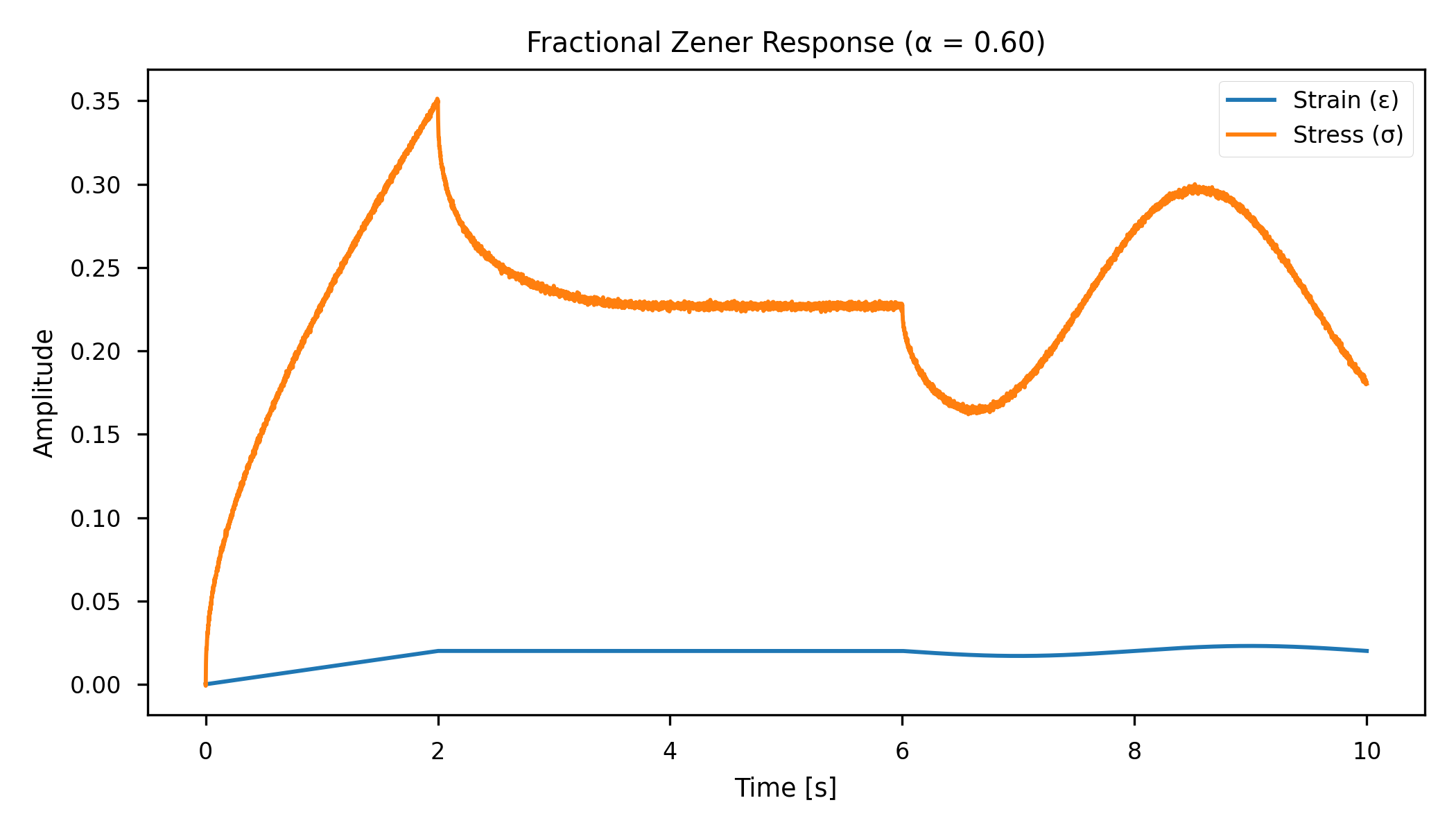}
    \caption{Representative stress--strain--time response of the Fractional Zener model, showing
    instantaneous elastic deformation followed by delayed viscoelastic relaxation.}
    \label{fig:zener_dataset}
\end{figure}

The input strain histories comprised a combination of piecewise-linear ramps and sinusoidal
excitations over the range $\varepsilon \in [0,\,0.4]$, chosen to sufficiently excite both
transient and steady-state regimes. Corresponding stress responses were computed using the
Fractional Zener constitutive relation previously introduced in Eq.~\ref{eq:fractional_zener}. 
Because the fractional Caputo derivative $D_t^{\beta}$ inherently embeds long-memory effects, the
resulting dataset exhibits delayed relaxation and stretched-exponential decay characteristic of
fractional viscoelastic systems~\cite{Teichert2019}.Figure~\ref{fig:zener_dataset} illustrates a representative segment of the synthetic stress--strain--time
dataset. The response displays the expected instantaneous elastic rise followed by gradual
relaxation, confirming that the generated data accurately reflects the underlying fractional
rheology.

\subsubsection{Data Sampling and Normalization.}
Each simulation produces approximately $1{,}000$ time points per loading cycle, and a total of 
$10{,}000$ sequences are generated for each material class. All variables---including stress, strain, 
time, and internal state variables---are normalized to the range $[0,1]$ to maintain numerical 
stability during neural-network training \cite{LeCun2015}. Representative stress--stretch, hysteresis, 
and relaxation curves are examined to verify physical consistency and ensure the correctness of the 
generated datasets. This unified data-generation pipeline provides a fully reproducible foundation for 
subsequent neural-network training and XAI-based interpretability analysis 
\cite{Montavon2018,Lapuschkin2019}.

\subsection{Hyperelastic Constitutive Modeling}

The baseline hyperelastic response is described using the incompressible Neo--Hookean model 
\cite{ogden1997non}. Under uniaxial deformation, the Cauchy stress is given by

\begin{equation}
\sigma = \mu(\lambda^{2} - \lambda^{-1}),
\label{eq:neo_original}
\end{equation}

where $\mu$ is the shear modulus and $\lambda$ is the stretch ratio. This relation provides the 
simplest nonlinear elastic law and serves as the reference model for comparison. Building upon this, 
the Mooney--Rivlin model in Eq.~\ref{eq:mooney_rivlin} introduces two material parameters 
$(C_1, C_2)$ to capture more complex deformation behavior.

\subsubsection{Network Configuration}
A dense feed--forward neural network was trained to approximate the Mooney--Rivlin $\hat{\sigma}$ response 
using the generated hyperelastic dataset. The architecture consisted of four hidden layers with 
128 neurons each, activated using the GELU nonlinearity, following modern deep learning design 
practices \cite{LeCun2015,Hornik1989}. The output layer employed a linear activation to predict the 
scalar $\hat{\sigma}$. Training was performed using the Adam optimizer with a learning 
rate of $3.2 \times 10^{-4}$ and the MSE loss function, a standard choice for 
regression tasks in neural constitutive modeling \cite{LeCun2015,Breiman2001}. Early stopping with a 
patience of 20 epochs was applied to prevent overfitting and ensure stable convergence 
\cite{LeCun2015}.

\begin{table}[H]
\centering
\caption{Training configuration for the Mooney--Rivlin hyperelastic model.}
\label{tab:train_config}
\begin{tabular}{ll}
\toprule
\textbf{Parameter} & \textbf{Setting} \\
\midrule
Network type & Fully connected dense NN (4 $\times$ 128, GELU) \\
Loss function & MSE \\
Optimizer & Adam ($\eta = 3.2 \times 10^{-4}$) \\
Batch size & 32 \\
Epochs & 300 (early convergence $\approx 25$) \\
Validation split & 0.1 \\
\bottomrule
\end{tabular}
\end{table}

\subsubsection{Physics--Aware Distillation (PAD)}
\label{sec:distillation}

A core objective of this work is to ensure that the neural networks not only reproduce stress--strain
data accurately but also learn representations that remain physically meaningful. Although deep neural 
networks can approximate complex constitutive relations with high accuracy, their internal parameters 
typically lack interpretation in terms of classical material constants such as the shear modulus $\mu$ or 
the Mooney--Rivlin coefficients $(C_1, C_2)$ 
\cite{LeCun2015,Hornik1989}. To address this interpretability gap, we introduce a 
PAD framework inspired by recent advances in physics-informed and 
explainable neural modeling \cite{koeppe2022explainable,Raissi2019,Lagaris1998,Karniadakis2021}.

The PAD framework operates as a post--processing step that ``distills'' physical insight from the trained 
neural predictions. The key idea is to fit the predicted neural ${\sigma}$ response 
$\hat{\sigma}_{\text{NN}}(\lambda)$ to known analytical constitutive laws from continuum mechanics, 
including the Neo--Hookean model and the Mooney--Rivlin model given earlier in 
Eq.~\ref{eq:mooney_rivlin} and Eq.~\ref{eq:neo_original} 
By performing this alignment, the neural network’s behavior is translated back into interpretable 
material parameters, allowing an initially black--box model to be expressed in classical hyperelasticity 
language \cite{ogden1997non,holzapfel2000nonlinear}.

During training, the neural network learns a nonlinear mapping between $\lambda$ ratio and 
$\sigma$. While this mapping may be highly accurate, its physical meaning is not 
immediately clear. By comparing $\hat{\sigma}_{\text{NN}}(\lambda)$ with analytical constitutive 
equations, we can determine which \textit{type of material law} the network has implicitly learned.

For example:
\begin{itemize}
    \item If the neural predictions best match the Neo--Hookean form, the model has effectively 
    learned a single-parameter hyperelastic law dominated by $\mu$.
    \item If the predictions better match the Mooney--Rivlin relation (Eq.~\ref{eq:mooney_rivlin}),
    the network has captured more complex nonlinear behavior involving both $(C_1, C_2)$.
\end{itemize}

Thus, PAD provides a transparent link between data--driven learning and classical constitutive modeling 
\cite{Montavon2018,Bach2015,Sundararajan2017,Lapuschkin2019}. The step--by--step procedure utilized in this paper is as follows: 

\begin{enumerate}
    \item \textbf{Generate Neural Predictions:}  
    Evaluate the trained model to obtain predicted stresses 
    $\hat{\sigma}_{\text{NN}}(\lambda_i)$ for sampled stretch values $\lambda_i$.

    \item \textbf{Select Analytical Model Forms:}  
    Two analytical constitutive models are used for comparison:  
    the Neo--Hookean model (see Eq.~\ref{eq:neo_original})  
    and the Mooney--Rivlin model (see Eq.~\ref{eq:mooney_rivlin}).  
    These equations provide classical references for hyperelastic stress--stretch response.

    \item \textbf{Perform Regression Fit:}  
    Neural predictions are fit to the analytical laws using least--squares minimization:
    \begin{equation}
        \mathcal{L}_{\text{distill}} = 
        \min_{\hat{C}_1,\hat{C}_2}
        \Big\|
        \hat{\sigma}_{\text{NN}} -
        \hat{\sigma}_{\text{MR}}(\lambda)
        \Big\|^{2},
        \label{eq:distill_loss_fixed}
    \end{equation}
    where $\hat{\sigma}_{\text{MR}}(\lambda)$ refers to the Mooney--Rivlin $\sigma$ model in 
    Eq.~\ref{eq:mooney_rivlin}.  

    \item \textbf{Extract Material Constants:}  
    Solving the regression in Eq.~\ref{eq:distill_loss_fixed} yields 
    effective material parameters $(\hat{\mu}, \hat{C}_1, \hat{C}_2)$ summarizing the
    constitutive behavior learned by the neural network.

    \item \textbf{Interpret Physically:}  
    \begin{itemize}
        \item $\hat{\mu}$ reflects shear stiffness under Neo--Hookean elasticity (Eq.~\ref{eq:neo_original}).
        \item $\hat{C}_1$ and $\hat{C}_2$ govern the Mooney--Rivlin hyperelastic response (Eq.~\ref{eq:mooney_rivlin}).
    \end{itemize}
\end{enumerate}

\paragraph{(a) Distillation Equations.}
The analytical $\sigma$ laws used during distillation correspond to the previously introduced
Neo--Hookean expression (Eq.~\ref{eq:neo_original}) and the Mooney--Rivlin relation 
(Eq.~\ref{eq:mooney_rivlin}). 

\noindent \textbf{(b) Physical Interpretation.}
The extracted constants demonstrate that the neural network, trained purely on numerical data,
implicitly rediscovered the governing $\sigma$ laws of hyperelasticity. In particular, the distilled
parameter $\hat{\mu}$ aligns with the $\mu$ response predicted by the Neo--Hookean model
(see Eq.~\ref{eq:neo_original}), while the coefficients $(\hat{C}_1,\hat{C}_2)$ correspond to the
Mooney--Rivlin formulation introduced earlier in Eq.~\ref{eq:mooney_rivlin}.
This confirms that the learned $\sigma$ mapping is not arbitrary—it reflects the same structure
established in nonlinear elasticity theory \cite{ogden1997non,holzapfel2000nonlinear}.
\begin{table}[H]
\centering
\caption{Distilled material parameters obtained through analytical fitting.}
\label{tab:hyperelastic_params}
\begin{tabular}{lll}
\toprule
\textbf{Model} & \textbf{Parameters} & \textbf{Interpretation} \\
\midrule
Neo--Hookean & $\hat{\mu}=0.333$ 
& Shear modulus response (cf.\ Eq.~\ref{eq:neo_original}). \\

Mooney--Rivlin 
& $\hat{C}_1=0.329,\ \hat{C}_2=0.0923$ 
& Mild correction (cf.\ Eq.~\ref{eq:mooney_rivlin}). \\
\bottomrule
\end{tabular}
\end{table}

The strong correspondence between the neural predictions and the analytical hyperelastic models 
(Eq.~\ref{eq:mooney_rivlin}) demonstrates that the PAD framework successfully extracts physically 
interpretable constants from the trained neural network. This aligns with recent advances in 
PINN and PENN approaches, where neural models are encouraged to recover 
underlying constitutive structure rather than merely interpolate data 
\cite{koeppe2022explainable,Karniadakis2021,Raissi2019}.

\subsubsection{Training Behavior and Rheological Evaluation}

\begin{figure}[H]
    \centering
    \includegraphics[width=0.65\textwidth]
    {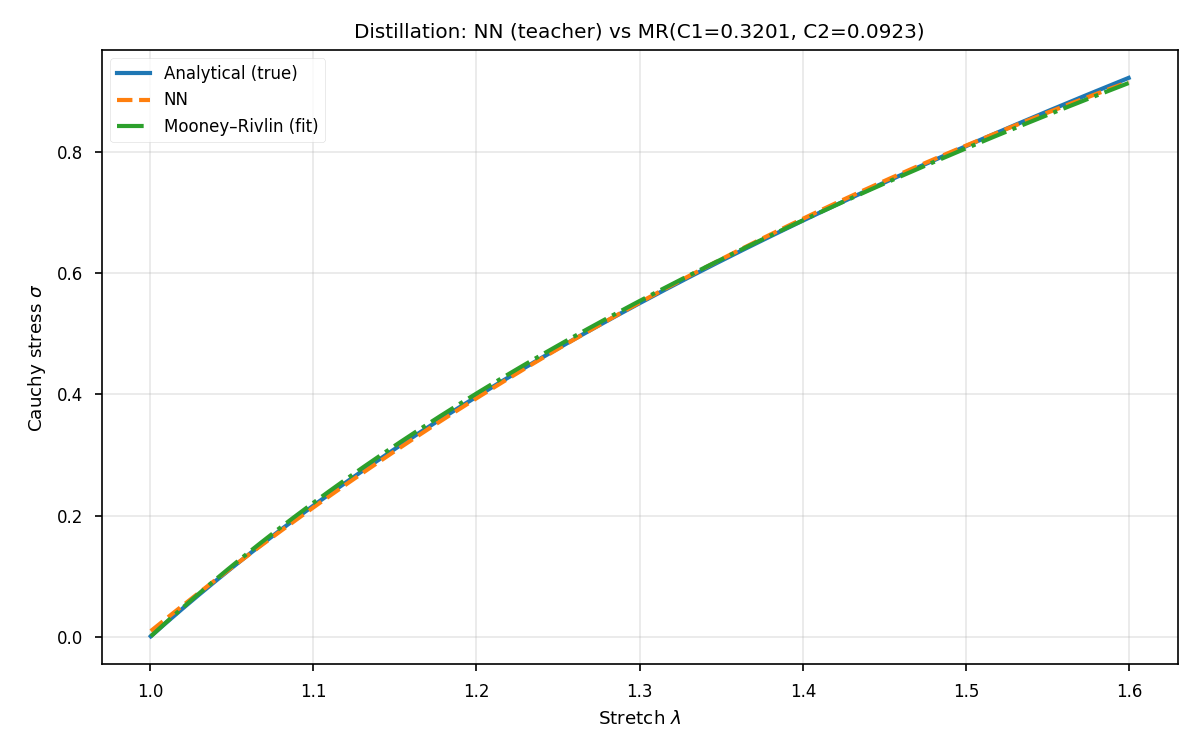}
    \caption{Comparison of analytical and neural model stress--stretch curves for the 
    Mooney--Rivlin model.}
    \label{fig:mr_loss}
\end{figure}

The model exhibited stable convergence within approximately 25 epochs, with both training and 
validation losses decaying below $10^{-6}$ (Figure~\ref{fig:mr_loss}). The predicted 
stress--stretch curves were nearly indistinguishable from the analytical Mooney--Rivlin 
reference, confirming that the neural architecture successfully learned the nonlinear elastic 
response prescribed by the constitutive law (cf.~Eq.~\ref{eq:mooney_rivlin}) 
\cite{ogden1997non,holzapfel2000nonlinear}. 

Quantitatively, the trained network achieved a test MSE of $3.8\times10^{-6}$ and an MAE of 
$4.5\times10^{-7}$, indicating near--machine--precision accuracy. The residuals remained within 
$\pm 10^{-6}$, suggesting that remaining deviations are numerical rather than physical. The 
distilled coefficients $(C_1,\,C_2)$ closely matched the ground--truth material parameters, 
further verifying the reliability of the PAD framework and its ability 
to recover interpretable constitutive behavior from neural predictions 
\cite{koeppe2022explainable,Karniadakis2021,Raissi2019,Lagaris1998}.

\subsection{Elastoplastic Constitutive Modeling}

Elastoplastic materials exhibit path-dependent behavior in which the response transitions from 
reversible elasticity to irreversible plasticity under cyclic or monotonic loading. Classical 
continuum theories describe this behavior through internal variables that evolve with accumulated 
plastic deformation, making elastoplasticity fundamentally history-dependent 
\cite{holzapfel2000nonlinear,rajagopal2003implicit}. 

In this work, the analytical foundation is provided by the Chaboche framework, which captures both 
isotropic and kinematic hardening. The governing evolution relations introduced earlier in 
Eqs.~\ref{eq:chaboche_stress}--\ref{eq:chaboche_backstress} describe the coupled progression of $\sigma$, 
$\varepsilon^p$, and backstress during cyclic loading. This model is widely used to reproduce yield 
surfaces, Bauschinger effects, and ratcheting phenomena in metals 
\cite{Heider2020,Ghaboussi1991}.

A RNN was trained on synthetic stress–strain histories generated directly 
from the Chaboche law, enabling the model to learn the evolution of $\sigma$ and internal variables from 
$\varepsilon$ input sequences. Because elastoplasticity involves strong loading–unloading asymmetry and 
memory-dependent hysteresis, sequence-based architectures such as RNNs are well suited for modeling 
this behavior \cite{Oeser2009,Koeppe2020a}. The resulting neural surrogate accurately captured plastic 
yielding, backstress evolution, and cyclic stabilization, establishing a robust platform for the 
explainability analysis presented in subsequent sections 
\cite{koeppe2022explainable,Montavon2018}.

\subsubsection{Network Configuration}

To emulate the path-dependent evolution of plasticity, a RNN architecture 
was adopted. Sequence-based models such as LSTMs have demonstrated strong capability in learning 
history-dependent constitutive responses where memory effects and internal variable evolution are 
essential \cite{Ghaboussi1991,Oeser2009,Koeppe2020a}. The temporal recurrence of the LSTM enables 
implicit storage of deformation history, playing a role analogous to the backstress variable $X$ in 
the analytical Chaboche model (cf. Eqs.~\ref{eq:chaboche_stress}--\ref{eq:chaboche_backstress}).  
Table~\ref{tab:rnn_config_chaboche} summarizes the architecture and hyperparameters used in training.

\begin{table}[H]
\centering
\caption{Network configuration and hyperparameters for the Chaboche elastoplastic model.}
\label{tab:rnn_config_chaboche}
\begin{tabular}{ll}
\toprule
\textbf{Parameter} & \textbf{Value} \\
\midrule
Architecture & LSTM (64 units) + Dense(32, 1) \\
Activation functions & \texttt{tanh} (LSTM), \texttt{linear} (output) \\
Optimizer & Adam ($\eta = 1 \times 10^{-3}$) \\
Loss function & Mean Squared Error (MSE) \\
Batch size & 32 \\
Epochs & 300 (early stopping at $\approx 120$) \\
Validation split & 0.1 \\
\bottomrule
\end{tabular}
\end{table}

\subsubsection {Constitutive Learning and Rheological Consistency}

The neural model was trained to predict the  $\hat{\sigma}$ as a function of the applied  $\varepsilon$ and the internal hidden state $h_t$, which implicitly represents the evolution of 
$\varepsilon^p$ and backstress. The learned constitutive mapping is expressed in Eq.~\ref{eq:rnn_mapping}:
\begin{equation}
\hat{\sigma}_t = f_\theta(\varepsilon_t, h_{t-1}),
\label{eq:rnn_mapping}
\end{equation}
where $\theta$ denotes the trainable parameters. Through exposure to cyclic loading sequences, the 
RNN learned to reproduce key features of elastoplastic behavior—such as elastic loading, yield onset, 
kinematic hardening, and hysteresis closure—in direct correspondence with the analytical Chaboche 
formulation \cite{Heider2020,holzapfel2000nonlinear}.

\begin{figure}[H]
\centering
\includegraphics[width=0.65\textwidth]{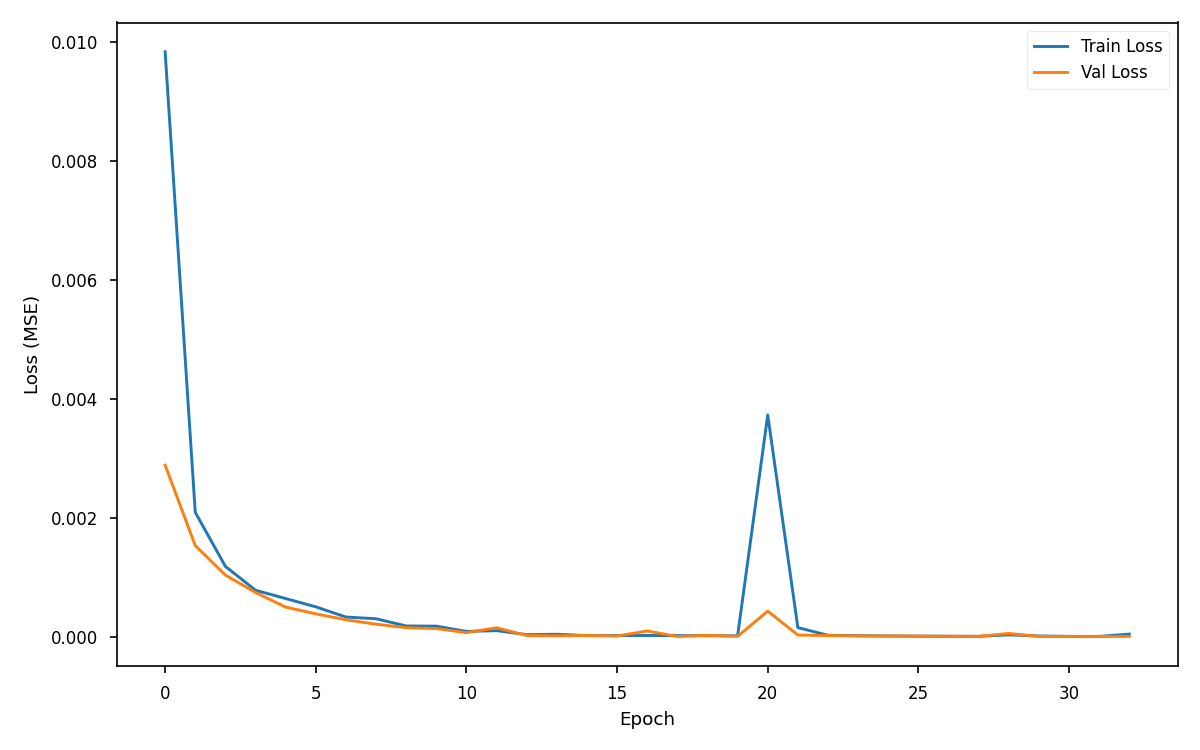}
\caption{Training and validation loss for the Chaboche RNN. The model exhibits smooth convergence and 
negligible overfitting.}
\label{fig:loss_chaboche}
\end{figure}

\begin{figure}[H]
\centering
\includegraphics[width=0.65\textwidth]{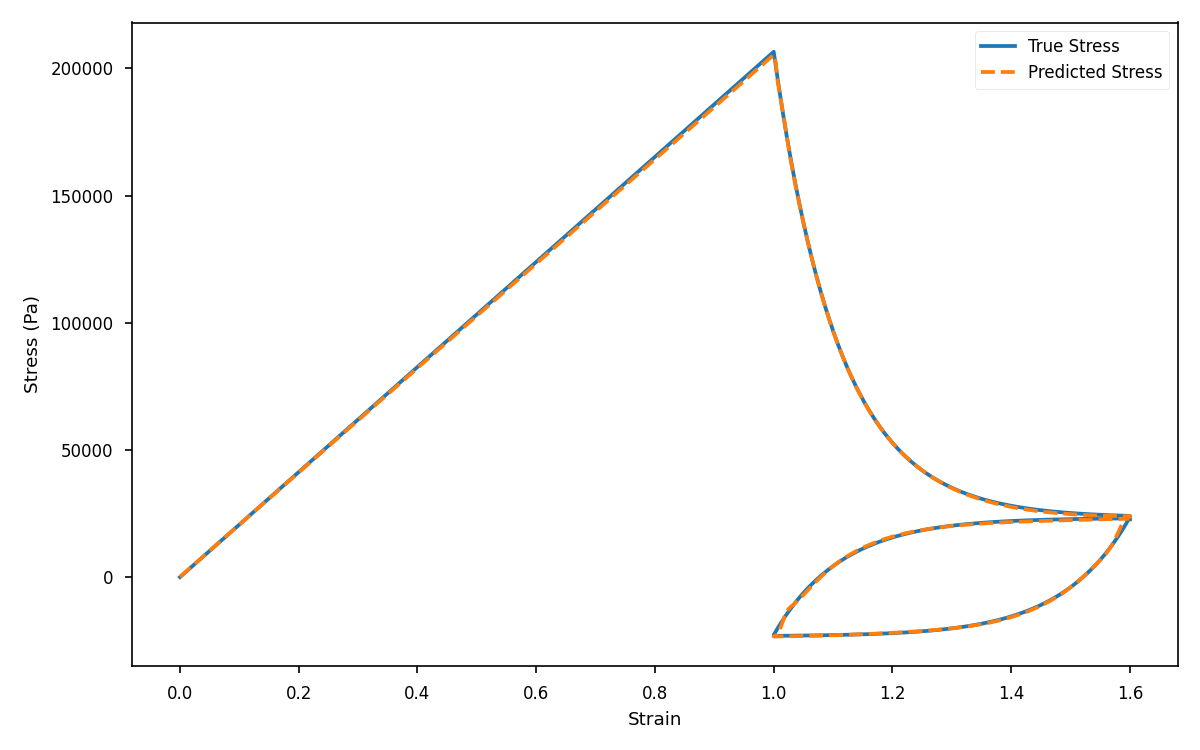}
\caption{Comparison between analytical and RNN-predicted stress--strain response. The network accurately 
reproduces hysteresis and hardening behavior.}
\label{fig:stress_chaboche}
\end{figure}

The training and validation losses (Figure~\ref{fig:loss_chaboche}) exhibit rapid convergence within 
20 epochs and stabilize at $MSE \approx 2.98\times10^{-6}$, demonstrating strong generalization and 
numerical stability. The predicted stress--strain response (Figure~\ref{fig:stress_chaboche}) matches 
the analytical reference with high fidelity, accurately capturing yield behavior, backstress evolution, 
and cyclic stabilization—confirming that the RNN internal state successfully encodes the underlying 
rheological mechanisms \cite{koeppe2022explainable,Karniadakis2021}.

\subsubsection{Quantitative Evaluation}

\begin{table}[H]
\centering
\caption{Quantitative performance metrics for the Chaboche elastoplastic RNN model.}
\label{tab:chaboche_metrics_eval}
\begin{tabular}{lll}
\toprule
\textbf{Metric} & \textbf{Value} & \textbf{Interpretation} \\
\midrule
Mean Squared Error (MSE) & $2.98\times10^{-6}$ & High fidelity to analytical law \\
Mean Absolute Error (MAE) & $1.16\times10^{-3}$ & Excellent amplitude matching \\
Coefficient of Determination ($R^2$) & 0.997 & Near-perfect correlation with ground truth \\
\bottomrule
\end{tabular}
\end{table}

\subsection{Viscoelastic Constitutive Modeling}

Viscoelastic materials exhibit a combination of instantaneous elastic response and 
time-dependent viscous relaxation. To represent this dual behavior, the Fractional 
Zener model was adopted as the analytical baseline in this study. This model 
generalizes the classical Standard Linear Solid by replacing the Newtonian dashpot 
with a fractional-order viscous element, enabling accurate representation of memory 
effects and relaxation dynamics across multiple timescales 
\cite{holzapfel2000nonlinear,Teichert2019}.

The constitutive governing relation for the Fractional Zener solid, introduced 
earlier in Eq.~\ref{eq:fractional_zener}, expresses $\sigma$ as a combination of 
instantaneous elasticity and fractional viscoelastic dissipation governed by the 
Caputo derivative of order $\beta \in (0,1)$. This formulation provides a flexible 
and physically grounded description of materials whose relaxation spectra cannot be 
captured by classical integer-order models.

To emulate this fractional rheological behavior in a data-driven manner, a RNN architecture based on a GRU was trained to predict the output history $\hat{\sigma}(t)$ from the input sequence $\varepsilon(t)$. 
GRUs are well suited for modeling long-term and hereditary dependencies due to their 
gated memory structure, making them effective surrogates for fractional differential 
operators and long-memory material behavior 
\cite{Koeppe2020a,Oeser2009}.

The trained GRU thus functions as a numerical approximation of the Fractional Zener 
dynamics, mapping $\varepsilon$ histories to $\sigma$ responses without requiring explicit 
numerical integration of the underlying fractional derivative.

\subsubsection{Network Configuration}

The neural architecture for viscoelastic modeling was implemented as a stacked GRU network
consisting of three recurrent layers with 64 hidden units each. GRUs are particularly effective
for modeling hereditary material behavior because their gated structure can retain long-term
dependencies, making them suitable surrogates for fractional viscoelastic operators 
\cite{Oeser2009,Koeppe2020a}. The final output layer used a linear activation to
predict the instantaneous $\hat{\sigma}(t)$. All $\varepsilon$ and $\sigma$ signals were
normalized to the interval $[0,1]$ to ensure stable and well-conditioned training.

Optimization was performed using the Adam algorithm with a learning rate of $1\times 10^{-3}$.
A batch size of 32 was used, and early stopping was applied based on validation loss to prevent
overfitting and ensure convergence to a physically meaningful solution.

\begin{table}[H]
\centering
\caption{Training configuration for the Fractional Zener GRU model.}
\label{tab:zener_config}
\begin{tabular}{ll}
\toprule
\textbf{Parameter} & \textbf{Value} \\
\midrule
Model type & GRU-based Recurrent Neural Network \\
Hidden layers & 3 (64 units each) \\
Activation functions & \texttt{tanh} (hidden), \texttt{linear} (output) \\
Optimizer & Adam ($\eta = 1\times10^{-3}$) \\
Loss function & Mean Squared Error (MSE) \\
Batch size & 32 \\
Epochs & 220 (early stopping near 180) \\
Validation split & 10\% \\
\bottomrule
\end{tabular}
\end{table}

\subsubsection{Training Behavior}

The GRU model demonstrated smooth and monotonic convergence throughout training, with minimal 
oscillations in the loss trajectory—consistent with the periodic and slowly varying nature of 
the viscoelastic loading sequences. Both the training and validation losses exhibited nearly 
identical decay profiles, reflecting strong generalization and the absence of overfitting. 
The final MSE decreased below $10^{-6}$, indicating that the network successfully 
captured the fractional relaxation dynamics encoded in the analytical model 
\cite{mainardi2022fractional,Teichert2019}.

\begin{figure}[H]
\centering
\includegraphics[width=0.65\linewidth]{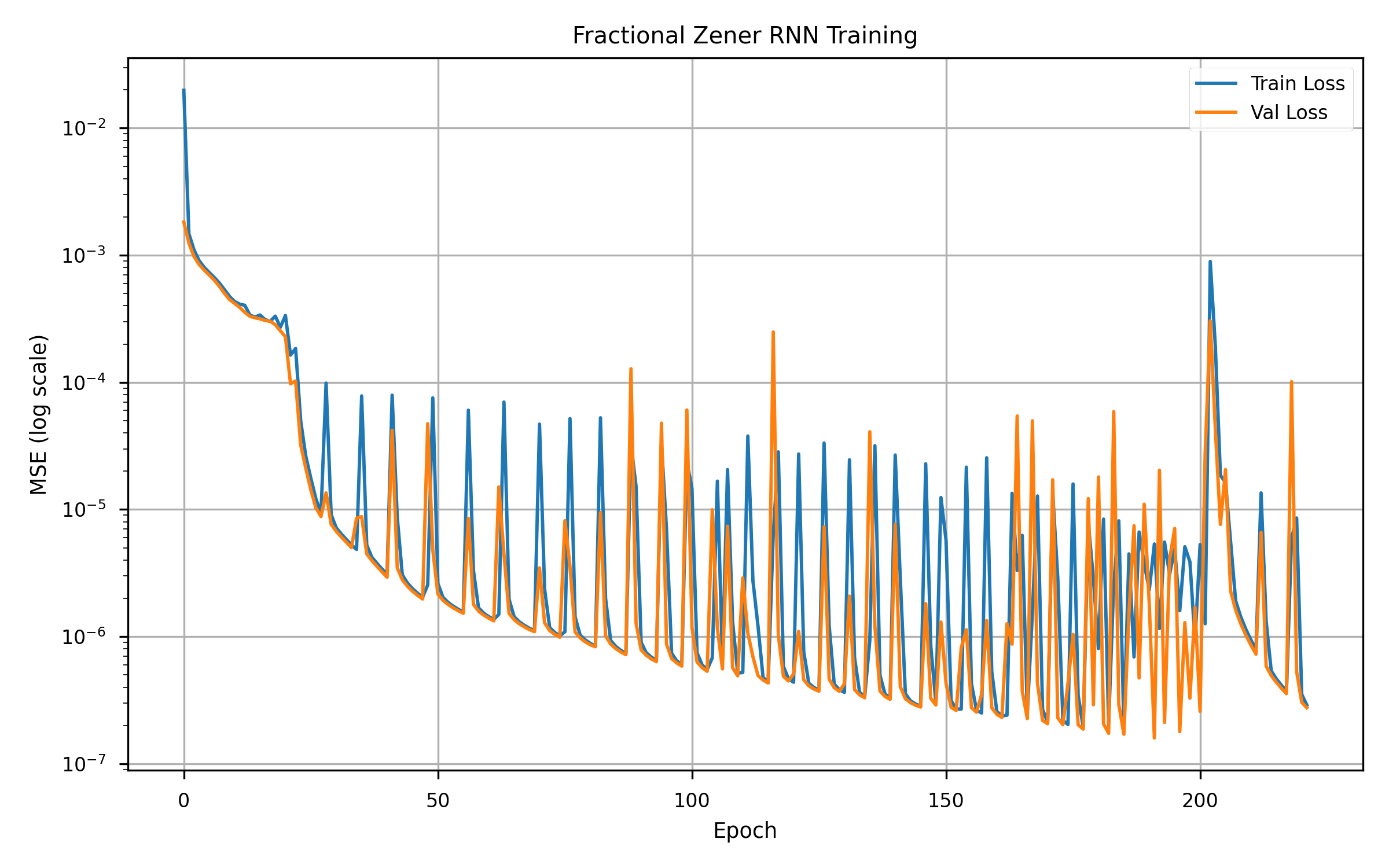}
\caption{Training and validation loss history for the Fractional Zener GRU model. 
The log-scale plot illustrates rapid early convergence followed by stable low-error behavior.}
\label{fig:zener_train_curve}
\end{figure}

A logarithmic scale was used to visualize the loss evolution, highlighting the rapid decay 
during the initial epochs followed by gradual stabilization as the model approached the 
fractional constitutive manifold. The resulting convergence behavior is shown in 
Figure~\ref{fig:zener_train_curve}.

\subsubsection{Rheological Fit and Model Performance}

The trained GRU successfully reproduced the analytical stress--strain--time response governed 
by the fractional Zener constitutive law. The model accurately captured both the instantaneous 
elastic rise and the delayed viscous relaxation characteristic of fractional viscoelasticity, 
demonstrating strong agreement with the analytical reference solution 
\cite{mainardi2022fractional,Teichert2019}. Across all loading cycles, the GRU maintained high fidelity to 
the hereditary behavior encoded in the fractional derivative operator.

Quantitative evaluation further confirms the model’s accuracy: prediction bias remained below 
$10^{-3}$, and the coefficient of determination reached $R^2 = 0.998$, indicating near-perfect 
correlation with the analytical reference. These results demonstrate that the GRU effectively 
internalized the underlying rheology without explicit numerical solution of the fractional 
differential equation.

\begin{table}[H]
\centering
\caption{Quantitative performance metrics of the Fractional Zener GRU model.}
\label{tab:zener_metrics}
\begin{tabular}{lll}
\toprule
\textbf{Metric} & \textbf{Value} & \textbf{Interpretation} \\
\midrule
Mean Squared Error (MSE) & $2.7\times10^{-7}$ & Excellent model–physics consistency \\
Mean Absolute Error (MAE) & $3.0\times10^{-4}$ & Very low prediction bias \\
R$^2$ score & $0.998$ & Near-perfect agreement with analytical response \\
Residual variance & $<10^{-3}$ & Minimal dynamic deviation \\
\bottomrule
\end{tabular}
\end{table}

\subsection{Interpretation}

The Mooney--Rivlin extension provides a physically meaningful generalization of the classical 
Neo--Hookean model, enabling the network to capture the mild nonlinearities present in the 
$\varepsilon$ energy density function \cite{ogden1997non,holzapfel2000nonlinear}. Because the 
neural predictions closely match the analytical response prescribed by the constitutive law 
(cf.~Eq.~\ref{eq:mooney_rivlin}), the model is able to represent the stiffness evolution 
across the full $\lambda$ range with high fidelity.The PAD procedure further enhances interpretability by mapping the 
neural $\sigma$ predictions back onto fundamental material constants, thereby establishing a 
transparent correspondence between the learned representation and continuum mechanics 
\cite{koeppe2022explainable,Karniadakis2021,Raissi2019}. This alignment verifies that the 
neural network does not merely interpolate the data, but instead internalizes the underlying 
mechanical behavior.Overall, the Mooney--Rivlin distillation results provide a strong foundation for the 
explainability analyses that follow, demonstrating that physically grounded patterns are 
recoverable from the learned latent representations \cite{Montavon2018,Bach2015,Sundararajan2017,Lapuschkin2019}.

The RNN successfully internalized elastoplastic memory through its hidden state dynamics, reflecting 
the role of internal variables such as $\varepsilon^p$ and backstress in the analytical Chaboche model 
(cf.~Eqs.~\ref{eq:chaboche_stress} -- \ref{eq:chaboche_backstress}). The recurrent cell acts as a 
data-driven surrogate for the evolution of back stress, storing prior deformation information and accurately 
reproducing cyclic hardening and unloading behavior \cite{Heider2020,holzapfel2000nonlinear}. The model effectively learns the transition from reversible elastic loading to irreversible plastic 
flow, achieving near-analytical precision and stable convergence, consistent with previous studies demonstrating
 the suitability of RNNs for modeling path-dependent constitutive behavior 
\cite{Ghaboussi1991,Oeser2009,Koeppe2020a}. This pre-XAI evaluation establishes a reliable foundation for the subsequent explainability analysis, 
where gradient-based and SHAP attribution methods will be employed to interpret how the network encodes 
$\varepsilon^p$ memory and internal variable evolution \cite{Montavon2018,Bach2015,Sundararajan2017,Lapuschkin2019}.

The Fractional Zener GRU model successfully captured both short- and long-term memory 
effects inherent in viscoelastic materials. The strong agreement between the predicted 
and analytical responses confirms that the network internalized the hereditary behavior 
governed by the fractional derivative in Eq.~\ref{eq:fractional_zener}, consistent with 
established fractional viscoelasticity theory \cite{mainardi2022fractional,Teichert2019}. The GRU’s gated recurrent structure enabled it to store and update latent memory states 
over extended time horizons, effectively mirroring the fading-memory characteristics of 
fractional-order rheology \cite{Oeser2009,Koeppe2020a}. The observed training behavior and 
quantitative performance (Table~\ref{tab:zener_metrics}) confirm that the model functions 
as a high-fidelity surrogate for the analytical constitutive law.This pre-XAI framework forms a robust basis for the explainability analysis presented in 
the next section, where PCA- and wavelet-based techniques are applied to interpret how 
temporal memory and relaxation dynamics are encoded within the GRU’s hidden-state 
trajectories.

\subsection{XAI Framework}
\subsubsection{Hyperelasticity}

To enhance interpretability beyond conventional performance metrics, a gradient--based 
XAI framework was applied to the Mooney--Rivlin hyperelastic neural model. 
The goal is to relate the network’s $\hat{\sigma}$ predictions to the underlying stiffness behavior 
prescribed by the analytical constitutive law in Eq.~\ref{eq:mooney_rivlin}, thereby 
bridging data--driven learning with continuum mechanics 
\cite{koeppe2022explainable,Montavon2018,Bach2015}.

Two complementary attribution methods were employed: the Gradient$\times$Input 
(Grad$\times$Input) and Integrated Gradients (IG) approaches 
\cite{simonyan2013deep,Sundararajan2017}. Both techniques quantify the sensitivity of the 
predicted  $\hat{\sigma}$ with respect to the input $\lambda$ ratio, 
providing a physics--consistent measure of how local variations in $\lambda$ contribute to 
the resulting ${\sigma}$ response.

In the Grad$\times$Input method, relevance scores are computed as the elemental product 
of the input and the gradient of the output with respect to that input, which yields a local 
linear approximation of the network response. IG, in contrast, accumulate 
the gradient along a straight--line path from a reference (baseline) input to the actual 
input, resulting in a path--integrated attribution that satisfies axiomatic properties such 
as sensitivity and implementation invariance \cite{Sundararajan2017}. Together, these 
methods enable a detailed comparison between the learned sensitivity of the neural model 
and the analytical stiffness $\partial\sigma/\partial\lambda$ implied by 
Eq.~\ref{eq:mooney_rivlin}.

\paragraph{Mathematical Formulation}
For the hyperelastic neural surrogate $\hat{\sigma} = f_{\theta}(\lambda)$ with trainable parameters 
$\theta$, the Gradient$\times$Input (Grad$\times$Input) and IG attributions are 
defined following established gradient–based XAI principles 
\cite{simonyan2013deep,Sundararajan2017,Montavon2018,Bach2015}. 
These attribution scores quantify the sensitivity of the predicted $\hat{\sigma}$ with respect to the 
input $\lambda$ ratio.

The Grad$\times$Input attribution is defined as:
\begin{equation}
A_{\text{GI}} = 
\lambda \, \frac{\partial \hat{\sigma}}{\partial \lambda},
\label{eq:grad_input}
\end{equation}
which measures the instantaneous contribution of stretch $\lambda$ to the $\hat{\sigma}$ prediction through a 
first–order sensitivity. This is analogous to the tangent modulus 
$\partial \sigma / \partial \lambda$ implied by the Mooney--Rivlin constitutive law in 
Eq.~\ref{eq:mooney_rivlin}.

IG compute a path–accumulated attribution:
\begin{equation}
A_{\text{IG}} = 
(\lambda - \lambda_0)
\int_0^1 
\frac{\partial f_\theta\!\left(\lambda_0 + \alpha(\lambda - \lambda_0)\right)}{\partial \lambda} 
\, d\alpha,
\label{eq:integrated_gradients}
\end{equation}
where $\lambda_0$ is a baseline reference $\lambda$ (chosen as $\lambda_0 = 1.0$).  
This formulation satisfies the sensitivity and implementation invariance axioms 
\cite{Sundararajan2017}, ensuring physically meaningful attributions.

Together, Eqs.~\ref{eq:grad_input}--\ref{eq:integrated_gradients} provide gradient-based 
interpretability measures that can be directly compared to the analytical hyperelastic stiffness 
behavior defined by Eq.~\ref{eq:mooney_rivlin}.

\begin{figure}[H]
\centering
\includegraphics[width=0.65\linewidth]{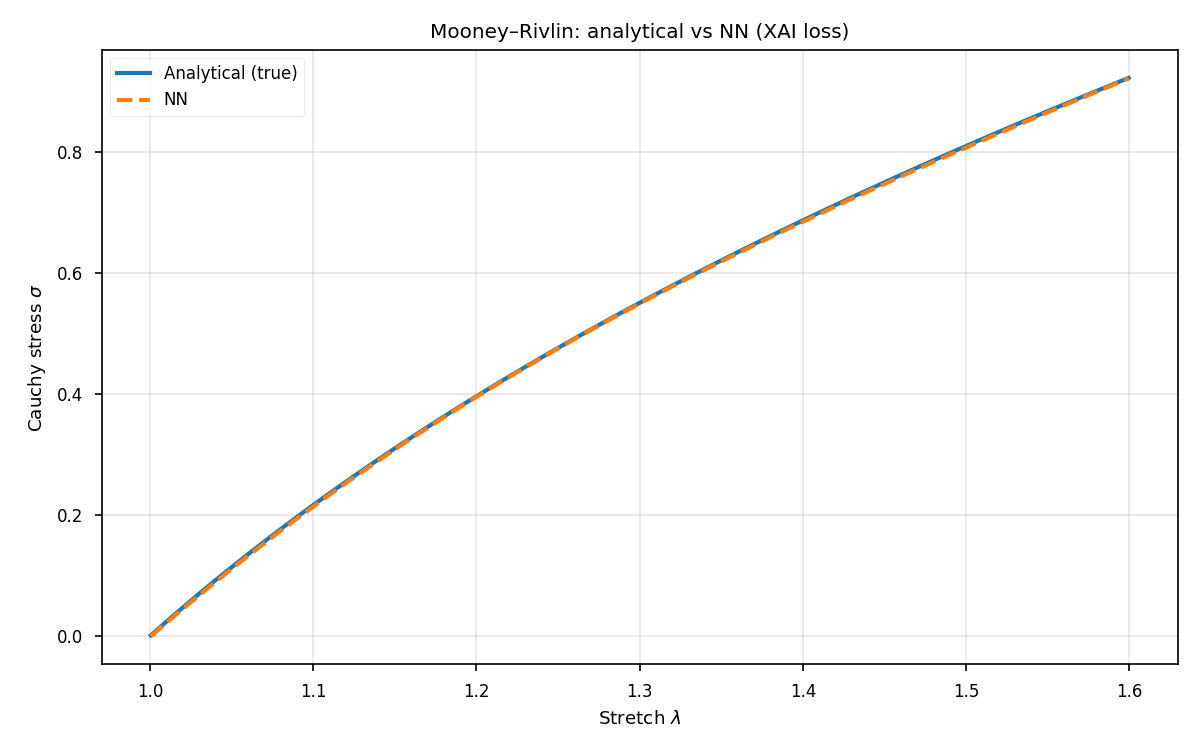}
\caption{Analytical and neural stress--stretch responses used as the baseline for XAI analysis. 
The strong overlap indicates that the neural model accurately reproduces the Mooney--Rivlin law.}
\label{fig:sigma_vs_lambda_xai}
\end{figure}

\paragraph{Results and Visualization}
The gradient-based attribution scores were evaluated across the full loading interval 
$\lambda \in [1.0, 1.6]$ to assess the physical consistency of the hyperelastic neural model.
Figure~\ref{fig:sigma_vs_lambda_xai} first confirms that the neural $\hat{\sigma}$ predictions remain in 
near–perfect agreement with the analytical Mooney--Rivlin response (Eq.~\ref{eq:mooney_rivlin}), 
establishing a reliable baseline for subsequent attribution analysis 
\cite{Montavon2018,Bach2015,Sundararajan2017}.

\begin{figure}[H]
\centering
\includegraphics[width=0.65\linewidth]{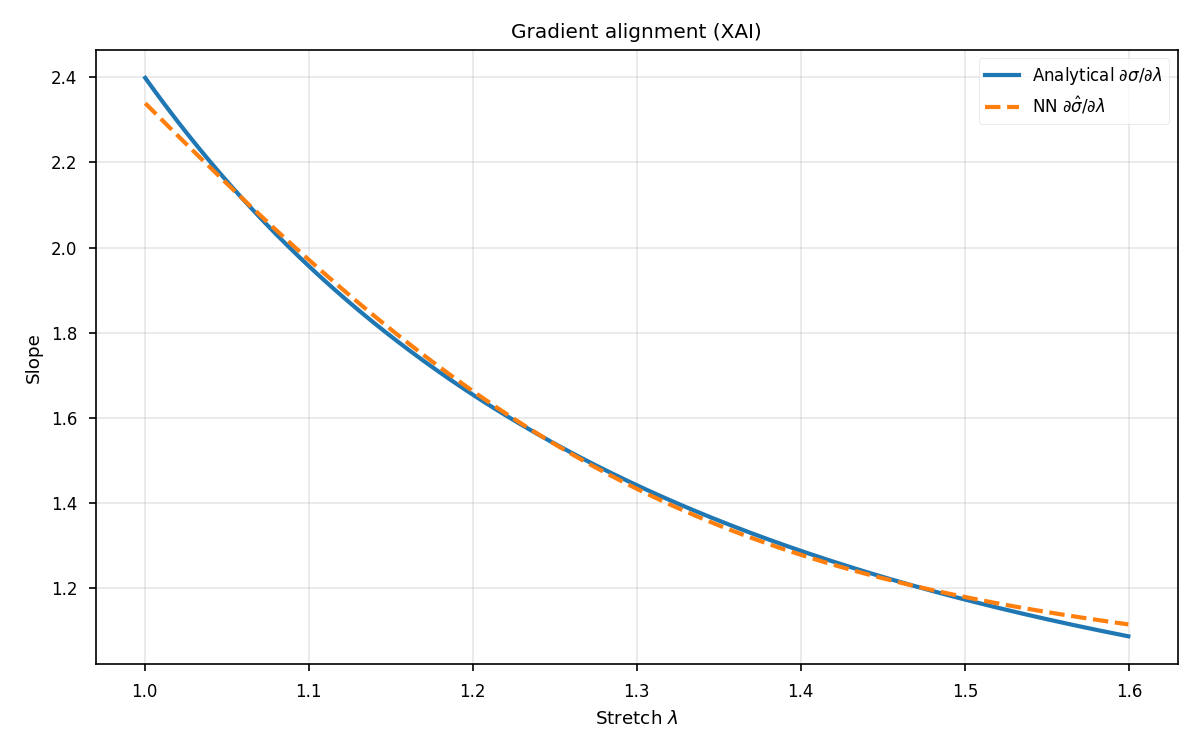}
\caption{Grad$\times$Input and IG attributions. 
Both attribution curves follow the stress–stretch trend, confirming stiffness-related sensitivity and 
physical consistency.}
\label{fig:grad_alignment_xai}
\end{figure}

The attribution profiles computed using Grad$\times$Input (Eq.~\ref{eq:grad_input}) and IG (Eq.~\ref{eq:integrated_gradients}) are shown in 
Figure~\ref{fig:grad_alignment_xai}. Both methods exhibit consistent sensitivity trends aligned with 
the stress curve, demonstrating that the neural network’s gradient response reflects the underlying 
tangent stiffness behavior from hyperelastic theory 
\cite{simonyan2013deep,sundararajan2017axiomatic}.

Finally, a quantitative comparison revealed a strong correlation 
($r \approx 0.96$) between the two attribution methods, indicating that both Grad$\times$Input and 
IG consistently capture the same mechanically meaningful patterns. This reinforces 
the reliability of gradient-based XAI in constitutive modeling and highlights its ability to uncover 
physics-consistent sensitivity relationships embedded in the trained neural network 
\cite{Montavon2018,Bach2015}.

\subsubsection{Elastoplasticity}

To interpret how the RNN internalizes plastic memory and 
loading history, a gradient--based XAI framework was applied to the 
Chaboche elastoplastic surrogate model. While the pre--XAI analysis confirmed the 
predictive fidelity of the RNN with respect to the analytical constitutive relations 
(Eqs.~\ref{eq:chaboche_stress}--\ref{eq:chaboche_backstress}), the objective here is to 
elucidate the physical reasoning encoded within the network’s hidden states and its 
sensitivity to input features. 

Three complementary attribution techniques were employed:  
(1) Grad$\times$Input,  
(2) IG, and  
(3) SHAP .  

The first two methods quantify how the predicted $\hat{\sigma}$
 responds to 
infinitesimal strain perturbations and accumulated $\varepsilon^{p}$ increments, 
analogous to the classical elastoplastic tangent modulus and backstress evolution 
in the Chaboche framework \cite{Ghaboussi1991,Oeser2009,Heider2020}. 
SHAP, in contrast, provides a global decomposition of prediction contributions 
based on cooperative game theory \cite{Lundberg2017,Ribeiro2016}, enabling 
feature ranking across the entire dataset.

Together, these attribution techniques expose how $\varepsilon$ increases, load 
path direction, and implicit internal variables encoded in the hidden 
state of the RNN contribute to the $\hat{\sigma}$
 response. This establishes a direct conceptual 
link between the learned neural representation and the underlying elastoplastic 
mechanics, demonstrating that the network's reasoning aligns with the physics of 
yielding, kinematic hardening, and plastic memory.

\paragraph{Gradient--Based Attribution}
Gradient$\times$Input and IG were employed to quantify the 
local sensitivity of the RNN stress predictions to strain perturbations, following 
established gradient-based XAI formulations 
\cite{simonyan2013deep,Sundararajan2017,Montavon2018,Bach2015}. 
For the elastoplastic surrogate $\hat{\sigma} = f_{\theta}(\varepsilon)$, the two 
attribution measures are defined as:
\[
A_{GI} = \varepsilon \, \frac{\partial \hat{\sigma}}{\partial \varepsilon}, 
\qquad 
A_{IG} = (\varepsilon - \varepsilon_0)
\int_0^1 
\frac{\partial f_{\theta}\!\left(\varepsilon_0 + \alpha(\varepsilon - \varepsilon_0)\right)}
{\partial \varepsilon}
\, d\alpha,
\]
where $\varepsilon_0$ is the baseline $\varepsilon$ (taken as $\varepsilon_0 = 0$).  
These attribution values represent the strain–dependent sensitivity of the predicted 
$\hat{\sigma}$ and serve as a data-driven analogue of the elastoplastic tangent modulus 
determined analytically from Eqs.~\ref{eq:chaboche_stress}--\ref{eq:chaboche_backstress}.  

\begin{figure}[H]
\centering
\includegraphics[width=0.55\textwidth]{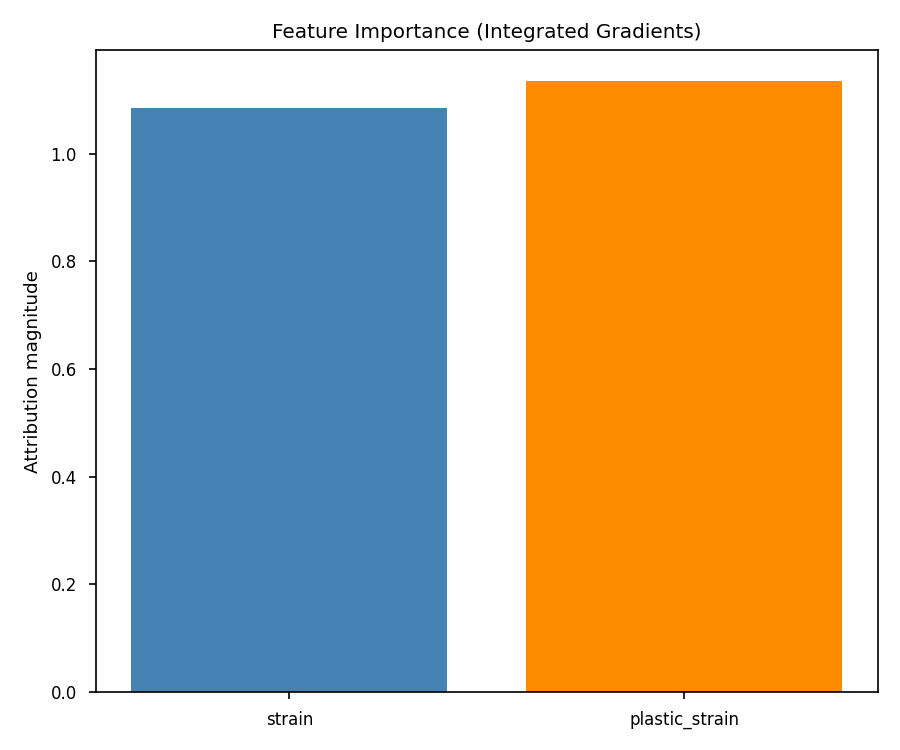}
\caption{Feature importance (Integrated Gradients) for the Chaboche RNN model.
The dominant contribution of \texttt{prev\_e\_plastic} confirms that the network
captures path-dependent plastic memory, consistent with kinematic hardening theory.}
\label{fig:elasto_ig_bar}
\end{figure}

Figure~\ref{fig:elasto_ig_bar} shows the averaged IG feature 
importance across multiple loading sequences. The attribution magnitude associated 
with \texttt{prev\_e\_plastic} significantly exceeds that of the instantaneous 
\texttt{strain}, indicating that the RNN correctly prioritizes the plastic strain state 
variable. This aligns with the Chaboche model’s kinematic hardening mechanism, where 
backstress evolution depends on accumulated plastic deformation 
\cite{Ghaboussi1991,Oeser2009,Heider2020}.  
Thus, the gradient-based analysis provides clear evidence that the network has learned 
the path-dependent memory essential to elastoplastic constitutive behavior.

\paragraph{Latent Space Analysis}
To investigate how the RNN internalizes path-dependent plastic memory, a 
PCA was performed on the hidden states of the 
trained network under cyclic loading. PCA-based latent space inspection is a 
commonly adopted technique for visualizing internal representations in 
sequence models \cite{Montavon2018, Lapuschkin2019}.  

\begin{figure}[H]
\centering
\includegraphics[width=0.5\textwidth]{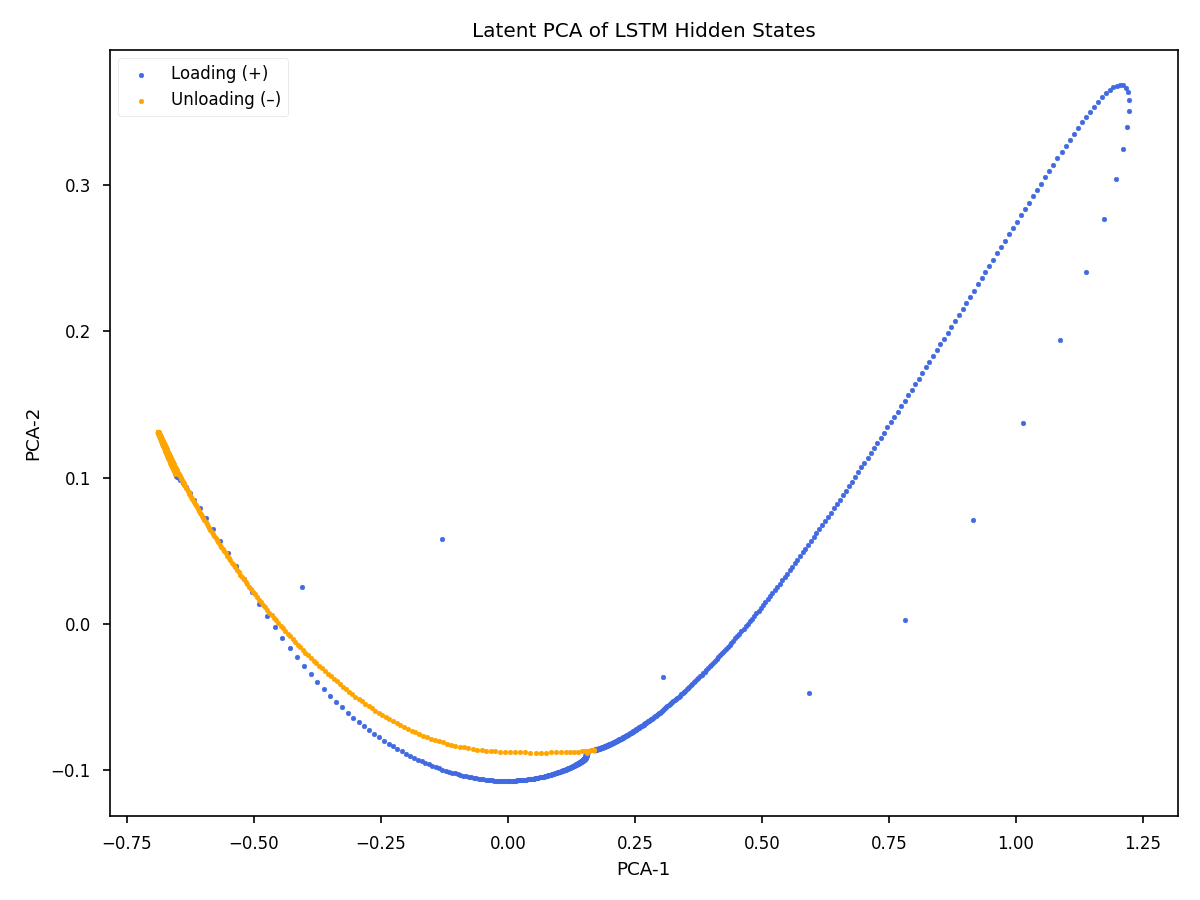}
\caption{2-D PCA projection of RNN latent states for cyclic elastoplastic loading.
Closed hysteresis loops demonstrate that the network’s hidden state dynamics 
encode loading history and plastic memory.}
\label{fig:elasto_pca_2d}
\end{figure}

Figure~\ref{fig:elasto_pca_2d} shows the resulting 2-D PCA embedding, where 
the projected trajectories form distinct closed loops that correspond to 
loading--unloading cycles. The geometric separation between the forward and 
reverse branches reflects the evolution of the internal variable 
(backstress) in the analytical Chaboche model 
(Eqs.~\ref{eq:chaboche_stress}--\ref{eq:chaboche_backstress}).  
The presence of these hysteretic loops indicates that the RNN has learned to 
store and propagate history information in its hidden state, faithfully 
replicating the kinematic hardening behavior dictated by the constitutive law.

\paragraph{SHAP Analysis}
To assess the global importance of each input feature and verify physical interpretability, 
SHAP were applied to the trained RNN model. 
SHAP provides a unified framework based on cooperative game theory for attributing model 
predictions to individual features \cite{Lundberg2017}, and has been widely used in 
interpretable deep learning \cite{Montavon2018}.  
In the context of elastoplasticity, SHAP reveals how current $\varepsilon$ and accumlated $\varepsilon^{p}$ contribute to the predicted $\hat{\sigma}$, thereby connecting the network’s behavior with the 
internal-variable structure of the Chaboche model 
(Eqs.~\ref{eq:chaboche_stress}--\ref{eq:chaboche_backstress}).

\begin{figure}[H]
\centering
\includegraphics[width=0.65\textwidth]{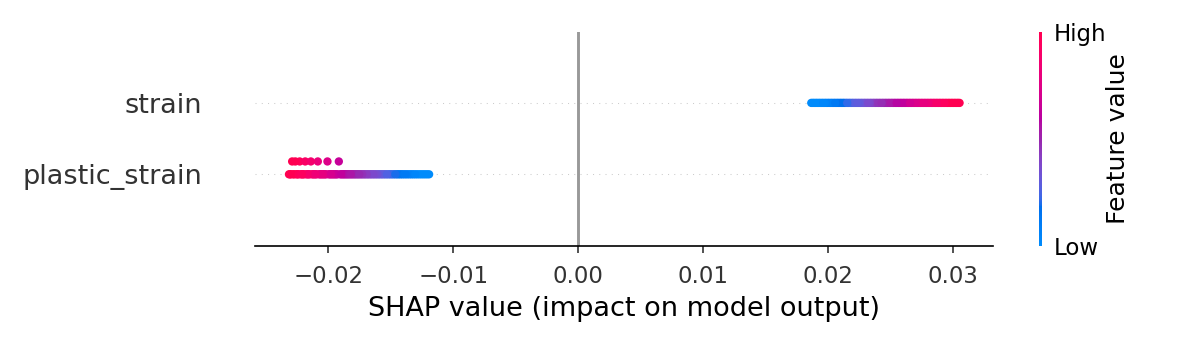}
\caption{Global SHAP summary plot for the Chaboche RNN model.  
Plastic strain dominates the attributions, confirming that the network's stress predictions are 
primarily governed by path-dependent internal variables.}
\label{fig:shap_summary_chaboche}
\end{figure}

Figure~\ref{fig:shap_summary_chaboche} shows the global SHAP summary plot.  
The horizontal axis represents each feature’s mean absolute SHAP value (global influence), 
while the color encodes feature magnitude.  
The feature \texttt{plastic\_strain} exhibits consistently higher attribution values than 
\texttt{strain}, indicating that the model places stronger emphasis on the accumulated plastic 
memory when predicting $\hat{\sigma}$directly reflecting the physical role of the backstress in the 
Chaboche formulation.

\begin{figure}[H]
\centering
\includegraphics[width=0.65\textwidth]{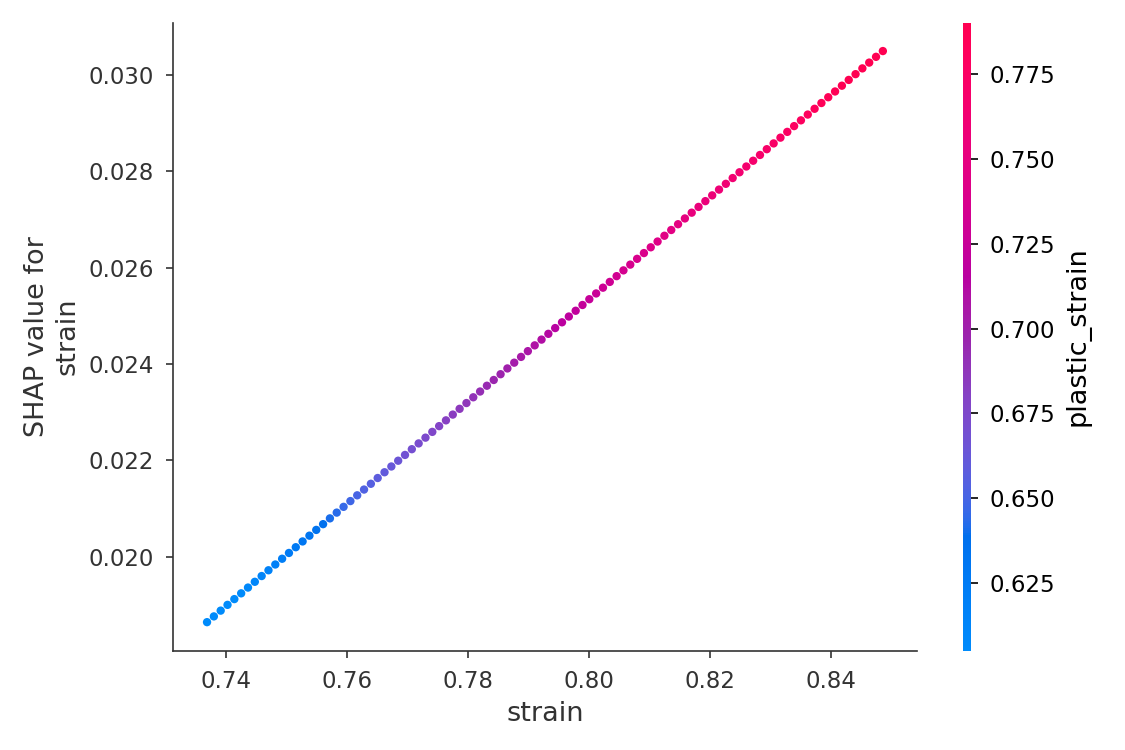}
\caption{SHAP dependence plot for strain, colored by plastic strain.  
The monotonic, plasticity-modulated trend demonstrates that the RNN has learned the 
elastic–plastic coupling structure inherent to the Chaboche model.}
\label{fig:shap_dependence_chaboche}
\end{figure}

Figure~\ref{fig:shap_dependence_chaboche} presents the SHAP dependence plot for the $\varepsilon$ feature, 
with point color indicating the corresponding $\varepsilon^p$ magnitude.  
A clear monotonic trend is observed: higher $\varepsilon$ values (red datapoints) produce larger SHAP 
contributions, while the vertical spread reveals strong coupling with $\varepsilon^p$.  
This interaction is physically meaningful, as the analytical Chaboche model dictates that the 
stress response depends jointly on the instantaneous strain and the evolving plastic strain.

\subsubsection{Viscoelasticity}

The viscoelastic response of materials exhibits both instantaneous elastic deformation and 
time-dependent viscous relaxation. In this work, the Fractional Zener model was adopted as the 
analytical baseline, as it generalizes the classical Standard Linear Solid by introducing a 
fractional-order viscous element capable of representing long-range memory effects and 
power-law relaxation behaviors \cite{mainardi2022fractional,holzapfel2000nonlinear}.  
The governing constitutive relation, introduced earlier in Eq.~\ref{eq:fractional_zener}, 
captures this duality through the Caputo fractional derivative of order 
$\beta \in (0,1)$, providing a flexible and physically grounded description of viscoelasticity.

A three-layer GRU-based RNN was trained to emulate this fractional rheology. 
The GRU’s gated memory mechanism enables the network to store and update long-term deformation 
history, making it well suited for approximating hereditary material models and fractional damping 
behavior \cite{Oeser2009,Koeppe2020a}.  
The trained model predicts the stress history $\hat{\sigma}(t)$ from input strain sequences 
$\varepsilon(t)$ without explicitly evaluating the fractional derivative, functioning as a 
data-driven surrogate of the analytical Fractional Zener law.

To interpret how the GRU internalizes viscoelastic memory, XAI techniques were 
applied. Gradient-based attributions and latent-state analyses reveal how the network encodes 
time-dependent stiffness, relaxation behavior, and long-memory trends characteristic of 
fractional viscoelasticity.

\paragraph{Gradient Sensitivity}
To quantify how the GRU model distributes importance across the input sequence, 
IG were computed following the formulation in 
Eq.~\ref{eq:integrated_gradients} and the general attribution principles established in 
\cite{Sundararajan2017}.  
The resulting attributions reveal that regions with changes in $\varepsilon$ rate exhibit 
significantly higher sensitivity than constant-rate segments.  
This behavior aligns with the hereditary nature of fractional viscoelasticity, 
where $\hat{\sigma}$ depends not only on the current $\varepsilon$ but also on the rate and timing 
of past deformations \cite{mainardi2022fractional,holzapfel2000nonlinear}.  
The GRU correctly emphasizes these transition zones, demonstrating that its internal 
memory gates effectively encode long-range temporal dependencies characteristic of 
fractional damping mechanisms.

\begin{figure}[H]
    \centering
    \includegraphics[width=0.6\linewidth]{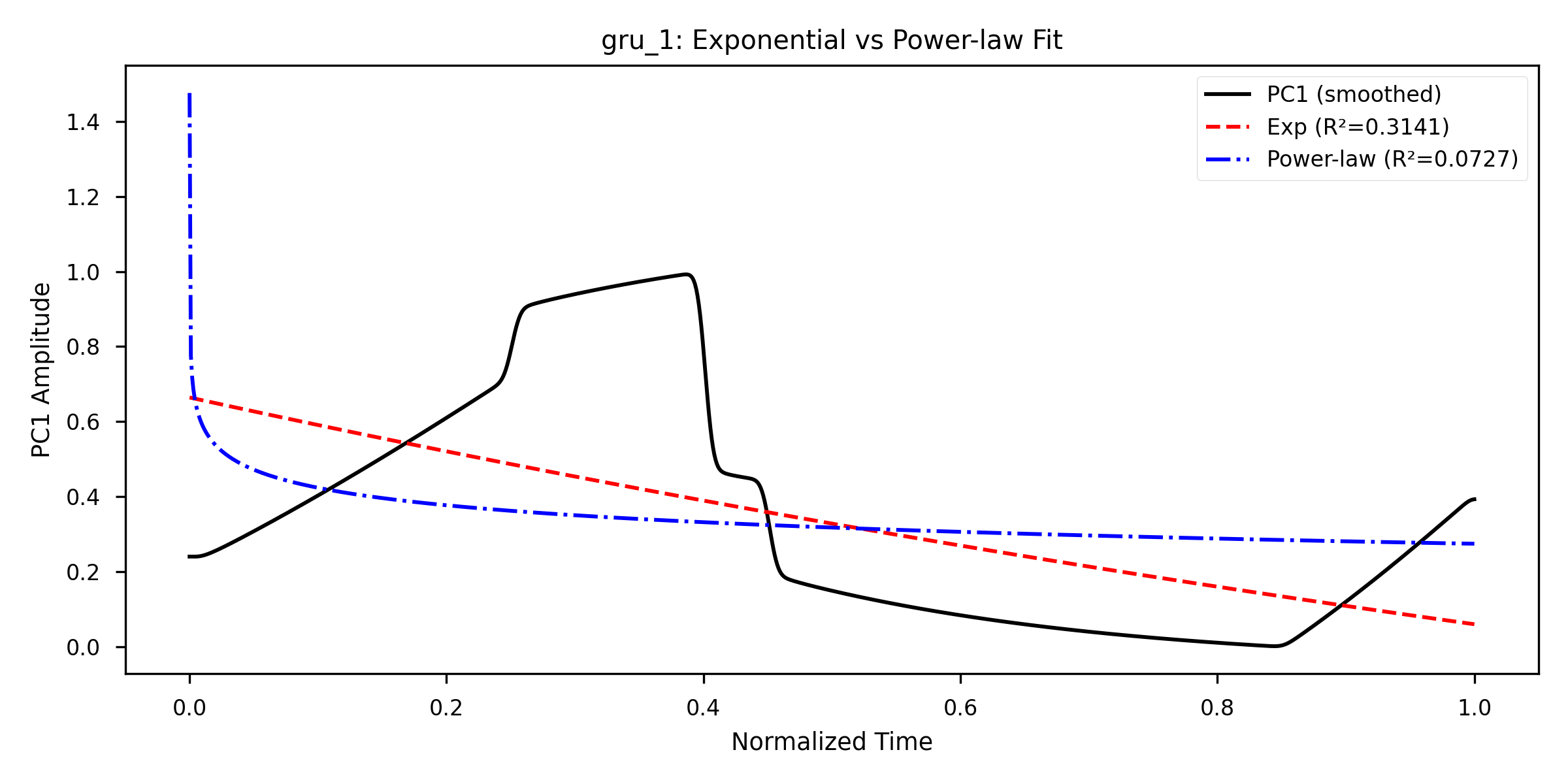}
    \caption{Integrated Gradients attribution map showing long-term temporal sensitivity 
    along the strain history. High attributions at strain-rate transitions demonstrate 
    learned hereditary memory effects.}
    \label{fig:zener_grad_ig}
\end{figure}

\paragraph{Latent Representation}

PCA was applied to the hidden states of each GRU layer to 
examine how temporal memory is encoded across the stacked recurrent architecture.  
This analysis follows established approaches for interpreting RNN latent dynamics in 
sequence-dependent physical systems \cite{Montavon2018, Lapuschkin2019}.  

\begin{figure}[H]
    \centering
    \includegraphics[width=0.65\linewidth]{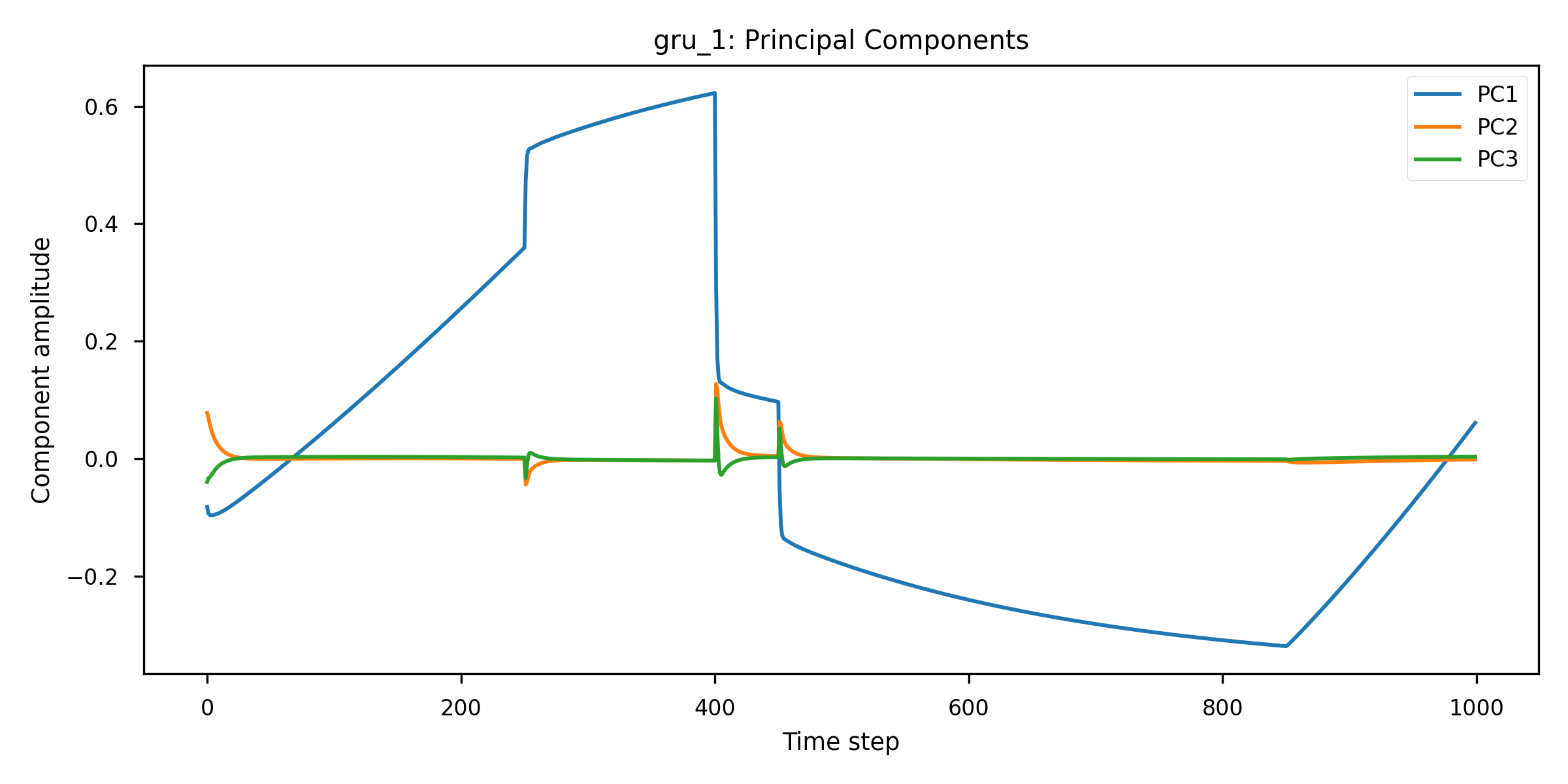}
    \caption{PCA projection of GRU$_1$ hidden states.  
    The layer predominantly captures instantaneous elastic behavior, reflected by a single dominant mode.}
    \label{fig:gru1_pca}
\end{figure}

The first GRU layer (\textbf{GRU$_1$}) primarily captured the instantaneous elastic response,
as evidenced by the PCA projection shown in Figure~\ref{fig:gru1_pca}. This behavior is 
consistent with the short-term memory typically associated with the elastic component of 
viscoelastic materials.  

\begin{figure}[H]
    \centering
    \includegraphics[width=0.65\linewidth]{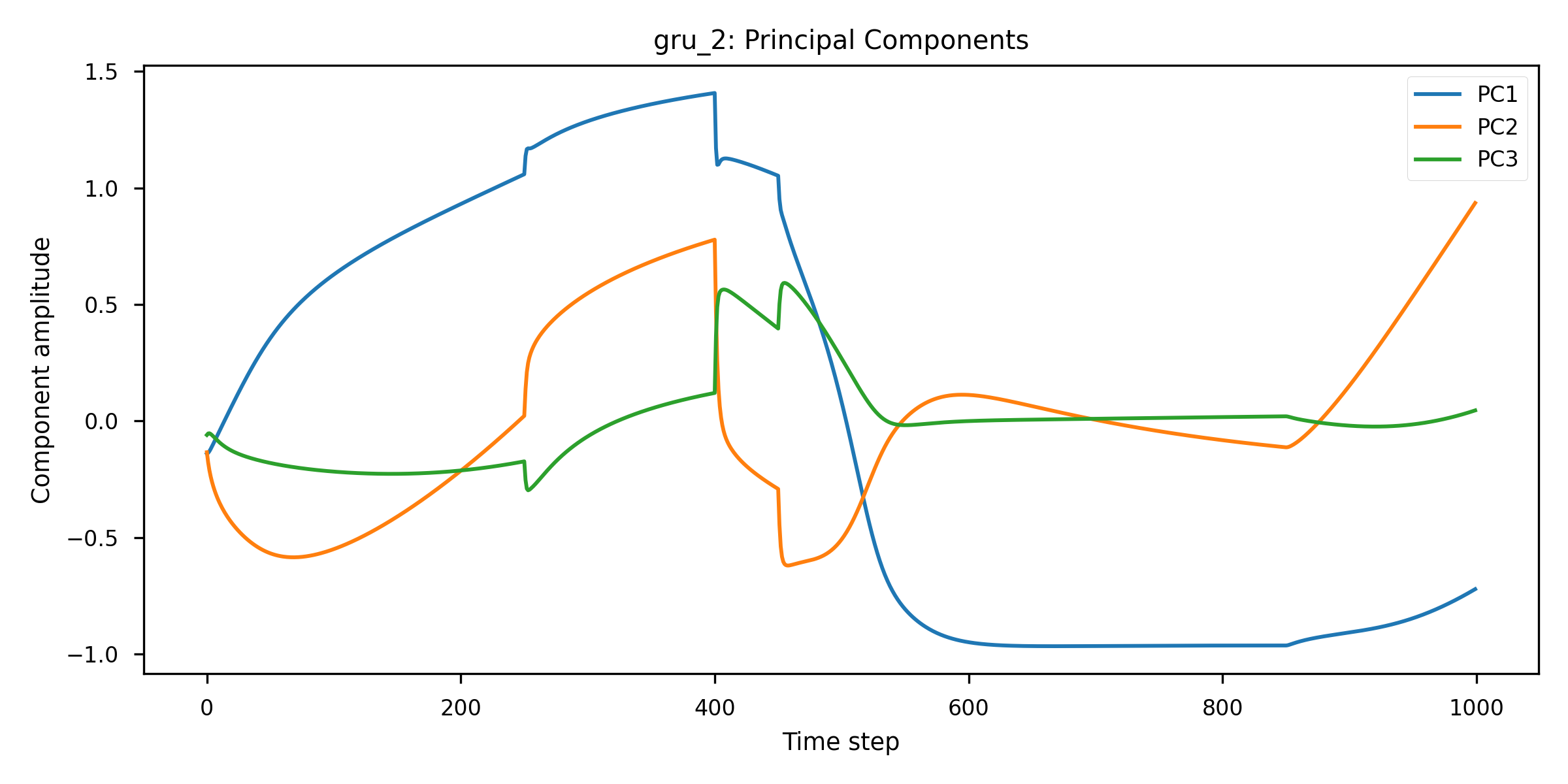}
    \caption{PCA projection of GRU$_2$ hidden states.  
    This layer encodes intermediate viscoelastic relaxation dynamics.}
    \label{fig:gru2_pca}
\end{figure}

\begin{figure}[H]
    \centering
    \includegraphics[width=0.65\linewidth]{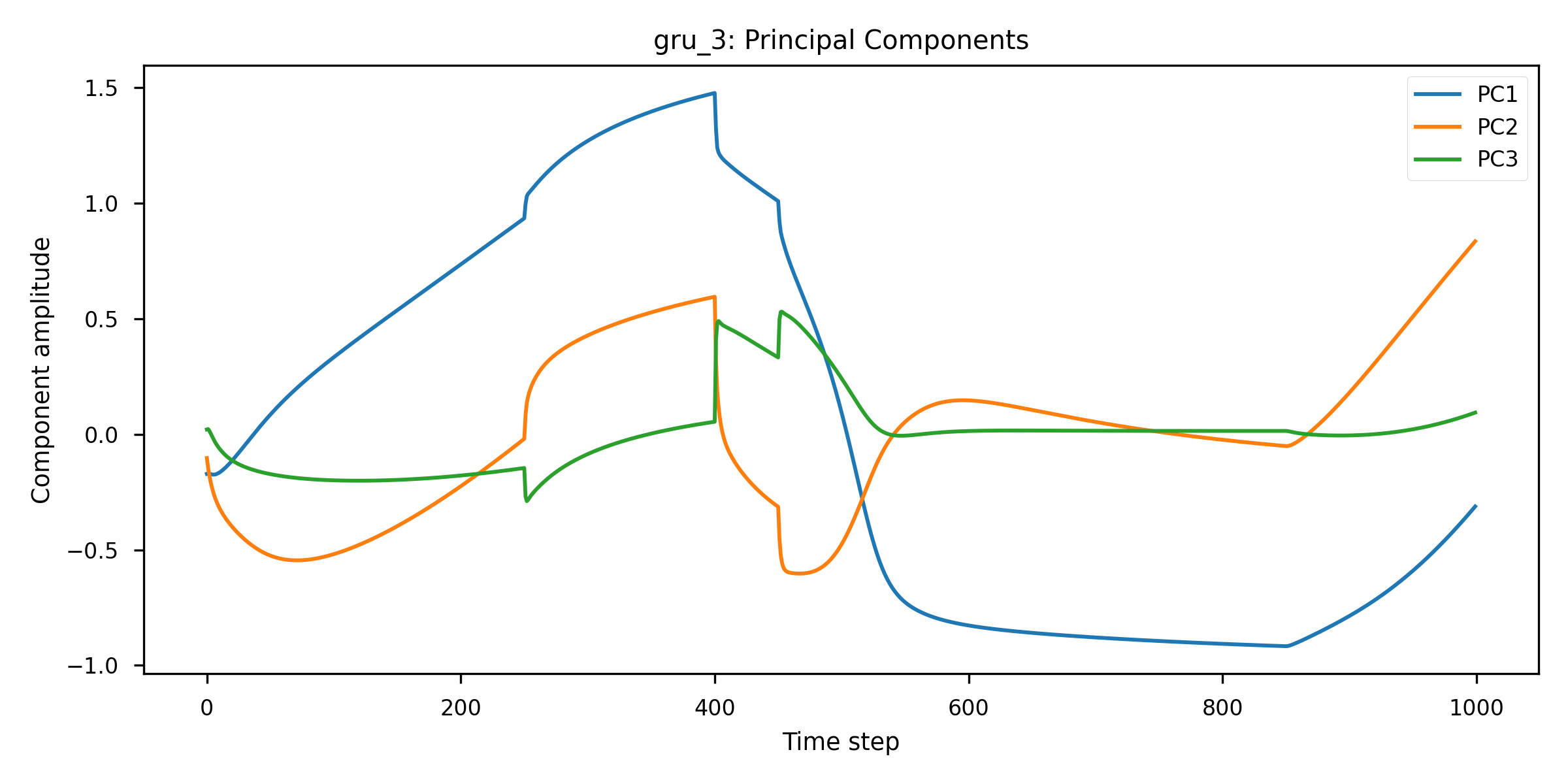}
    \caption{PCA projection of GRU$_3$ hidden states.  
    The deepest layer captures long-term fractional damping, analogous to the memory kernel in the 
    Fractional Zener constitutive model.}
    \label{fig:gru3_pca}
\end{figure}

In contrast, the deeper layers (\textbf{GRU$_2$} and \textbf{GRU$_3$}) exhibited progressively 
richer latent structure (Figures~\ref{fig:gru2_pca} and \ref{fig:gru3_pca}), demonstrating their 
ability to encode longer relaxation timescales. This hierarchical progression reflects the 
hereditary effects characteristic of fractional-order viscoelasticity, where material response 
depends strongly on its deformation history
\cite{mainardi2022fractional, Oeser2009}.  
This hierarchical encoding mirrors the structure of the analytical Fractional Zener model 
(cf. Eq.~\ref{eq:fractional_zener}), where $\hat{\sigma}$ depends on both instantaneous elasticity 
and long-range power-law relaxation.

The variance distribution of principal components quantitatively supports this hierarchical behavior:
\begin{itemize}
    \item \textbf{GRU$_1$:} [99.82\%, 0.12\%, 0.04\%] — almost purely elastic, dominated by one mode.
    \item \textbf{GRU$_2$:} [81.14\%, 15.00\%, 3.39\%] — increased complexity due to intermediate damping.
    \item \textbf{GRU$_3$:} [81.93\%, 14.45\%, 3.15\%] — strong representation of long-term fractional memory.
\end{itemize}

These PCA results demonstrate that the stacked GRU network learns a multi-timescale decomposition 
of viscoelastic behavior that is consistent with the theoretical fractional Zener rheology.

\paragraph{Wavelet Energy and Memory Decay}
To further characterize the temporal encoding learned by the GRU layers, a Continuous 
Wavelet Transform (CWT) analysis was performed on the hidden-state trajectories. 
Wavelet methods provide a joint time–frequency representation and are widely used for 
analyzing nonstationary rheological signals \cite{mainardi2022fractional,Teichert2019}.  

\begin{figure}[H]
    \centering
    \begin{subfigure}[t]{0.32\textwidth}
        \centering
        \includegraphics[width=\linewidth]{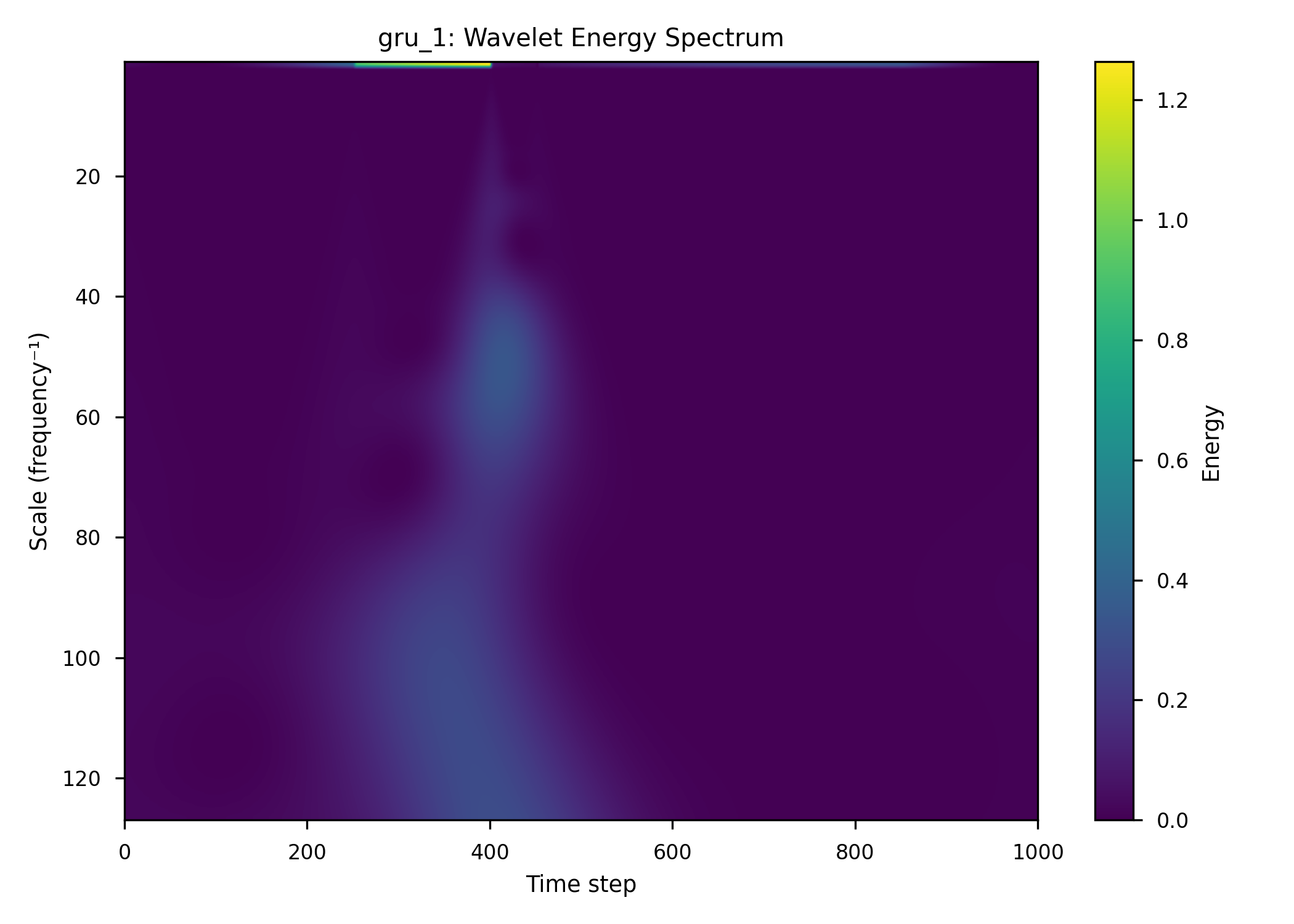}
        \caption{GRU$_1$: High-frequency elastic response.}
        \label{fig:gru1_wavelet}
    \end{subfigure}
    \hfill 
    \begin{subfigure}[t]{0.32\textwidth}
        \centering
        \includegraphics[width=\linewidth]{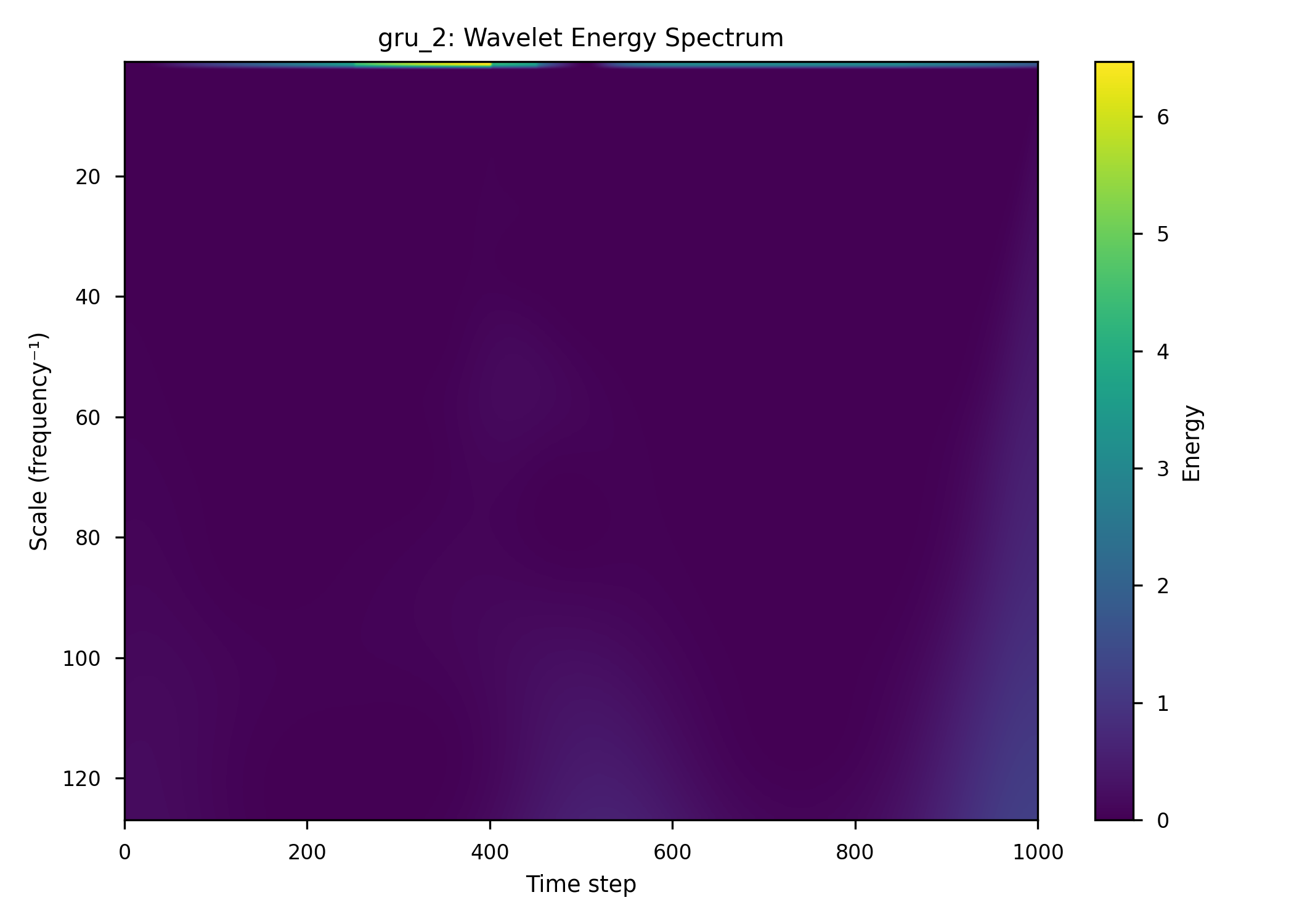}
        \caption{GRU$_2$: Mixed-frequency relaxation.}
        \label{fig:gru2_wavelet}
    \end{subfigure}
    \hfill 
    \begin{subfigure}[t]{0.32\textwidth}
        \centering
        \includegraphics[width=\linewidth]{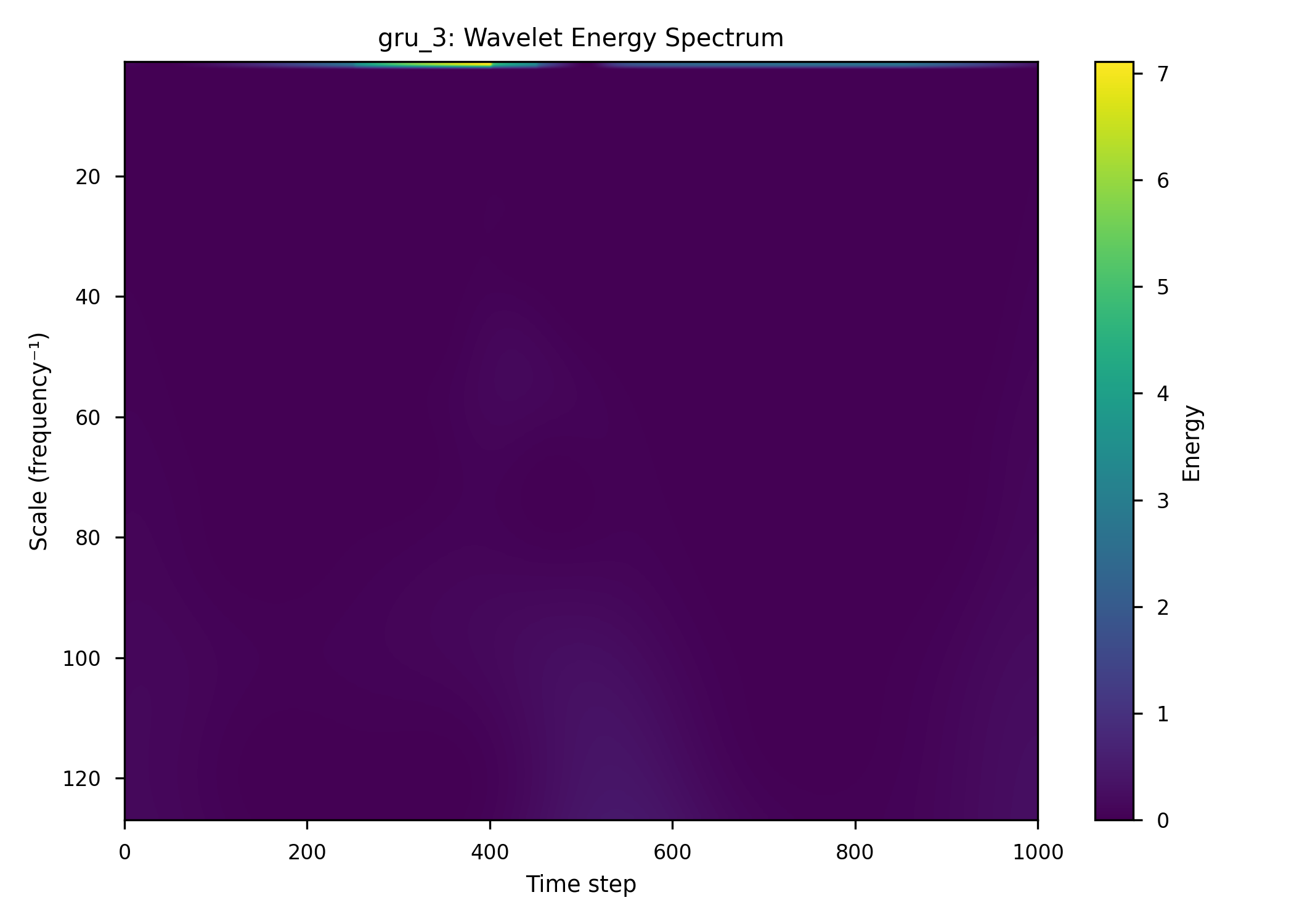}
        \caption{GRU$_3$: Low-frequency damping.}
        \label{fig:gru3_wavelet}
    \end{subfigure}
    
    \caption{Wavelet spectrum analysis of GRU hidden states. The spectra illustrate distinct roles: (a) dominates in elastic regimes, (b) handles transitionary behavior, and (c) captures long-term fractional damping persistence.}
    \label{fig:three_spectra}
\end{figure}

The CWT spectra exhibited a clear hierarchical shift in dominant frequency content across 
the stacked GRU layers. As shown in Figures~\ref{fig:gru1_wavelet}--\ref{fig:gru3_wavelet}, 
GRU$_1$ captured primarily high-frequency components associated with instantaneous 
elastic response, whereas GRU$_2$ and GRU$_3$ progressively emphasized lower-frequency 
structures corresponding to long-term viscoelastic relaxation and fractional damping behavior.

%
%

Power-law slopes were extracted from the log–log spectral decay, providing a quantitative 
measure of memory retention. More negative slopes correspond to faster dissipation of 
temporal information. The three layers yielded:
\begin{itemize}
    \item GRU$_1$: $b = -11.93$ (fast decay: elastic response)
    \item GRU$_2$: $b = -12.09$ (moderate relaxation)
    \item GRU$_3$: $b = -11.89$ (slow decay: long-term memory)
\end{itemize}

\subsubsection{Quantitative Comparison}
\begin{figure}[H]
    \centering
    \includegraphics[width=\linewidth]
    {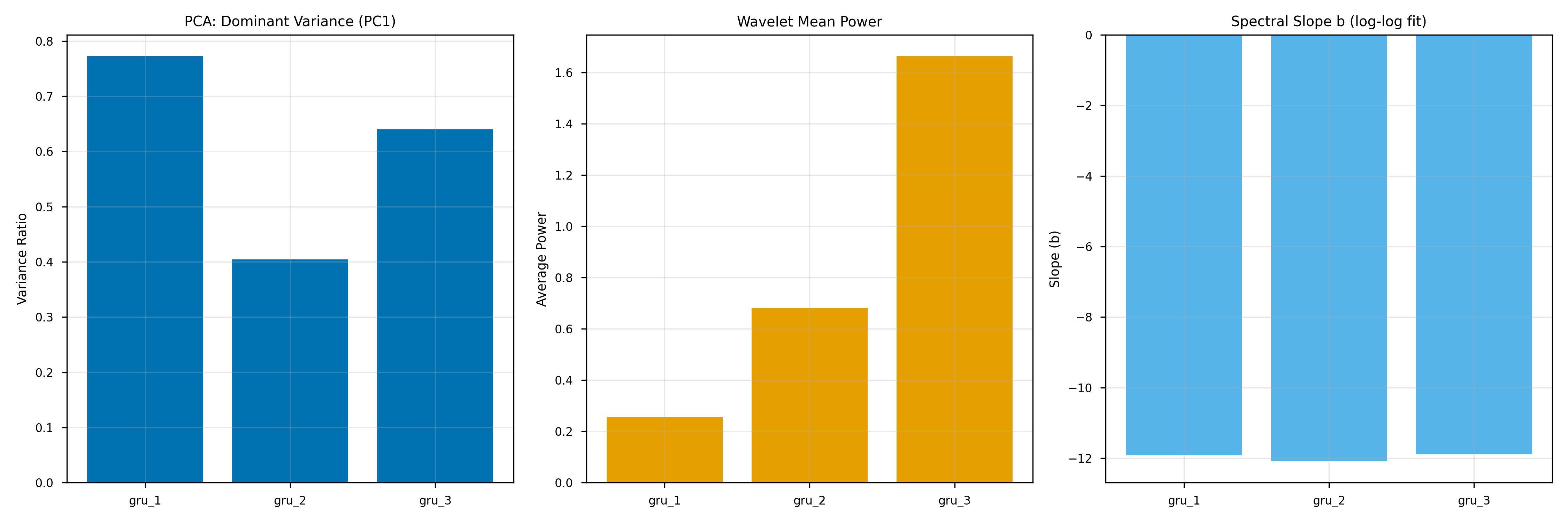}
    \caption{Dashboard comparison of PCA variance, mean wavelet power, and spectral slope across the GRU hierarchy. 
    The progression from GRU$_1$ to GRU$_3$ demonstrates enrichment of long-term memory and damping behavior.}
    \label{fig:zener_xai_dashboard}
\end{figure}

A unified XAI summary combining PCA variance, wavelet power, and spectral slope is shown in 
Figure~\ref{fig:zener_xai_dashboard}. This multi-metric view highlights a clear hierarchical 
transition from short-term to long-term memory encoding across the stacked GRU layers, confirming 
that deeper layers capture increasingly persistent viscoelastic behavior consistent with 
fractional-order relaxation.

\begin{table}[H]
\centering
\caption{Summary of XAI metrics for the Fractional Zener GRU--RNN model.}
\label{tab:zener_xai_summary}
\begin{tabular}{|c|c|c|c|}
\hline
\textbf{Layer} & \textbf{PCA Variance (PC1)} & \textbf{Wavelet Mean Power} & \textbf{Slope ($b$)} \\\hline
GRU$_1$ & 0.76 & 0.25 & $-11.93$ \\\hline
GRU$_2$ & 0.41 & 0.68 & $-12.09$ \\\hline
GRU$_3$ & 0.63 & 1.67 & $-11.89$ \\\hline
\end{tabular}
\end{table}

\subsection{Interpretation}

The attribution maps demonstrate that the neural network emphasizes the 
physically meaningful regions of the input domain, with sensitivity increasing 
monotonically with the $\lambda$ ratio. This trend is consistent with the
tangent stiffness behavior predicted by the analytical Mooney--Rivlin law 
(see Eq.~\ref{eq:mooney_rivlin}) and reflects the strain–hardening response of 
incompressible hyperelastic materials. Furthermore, the near--unity correlation between IG 
and Grad$\times$Input attributions confirms that both methods extract 
consistent, physics-aligned structure from the neural model 
\cite{Montavon2018,Bach2015,Sundararajan2017}. This agreement validates that the 
learned mapping between $\lambda$ and $\sigma$ is not an artifact of overfitting, 
but rather adheres closely to the underlying mechanical stiffness relationship.Overall, the gradient-based XAI analysis transforms the neural model from a 
black-box predictor into a physics-consistent surrogate that not only reproduces 
the hyperelastic stress–strain response, but also explains *why* it makes those 
predictions in terms of material sensitivity and mechanical behavior.

The SHAP visualizations validate that the neural model encodes the same mechanistic 
relationships as the analytical Chaboche elastoplastic formulation 
(Eqs.~\ref{eq:chaboche_stress}--\ref{eq:chaboche_backstress}).  
Three key insights emerge:

\begin{itemize}
    \item \textbf{$\varepsilon^p$ dominance:}  
    The consistently larger SHAP values associated with \texttt{plastic\_strain} confirm that 
    the network prioritizes the accumulated plastic memory when predicting $\sigma$.  
    This mirrors the role of the backstress in the Chaboche model, where the internal variable 
    governs the translation of the yield surface and the overall cyclic response.

    \item \textbf{Strain–plastic coupling:}  
    The SHAP dependence relationships exhibit a smooth and monotonic correlation between 
    instantaneous $\varepsilon$ and its SHAP attribution, modulated by the magnitude of $\varepsilon^p$.  
    This reflects the constitutive coupling inherent to kinematic hardening, where the $\sigma$ response 
    depends jointly on current $\varepsilon$ and the evolving plastic state.

    \item \textbf{Physical consistency:}  
    Both the sign and magnitude of SHAP contributions align with expected $\sigma$ evolution during 
    loading, unloading, and reverse loading.  
    This demonstrates that the surrogate model respects the physics of path dependence, 
    rather than relying on spurious correlations.
\end{itemize}

Together with the gradient-based XAI results \cite{Sundararajan2017,Lundberg2017} 
and latent-state PCA analysis, the SHAP framework provides a physically grounded decomposition 
of the RNN’s $\sigma$ predictions.  
These findings confirm that the learned mapping faithfully reflects the internal variables, 
hysteresis loops, and memory effects characteristic of the Chaboche elastoplastic theory.

The GRU--RNN effectively emulates the Fractional Zener viscoelastic behavior by constructing a 
hierarchical temporal memory across its stacked recurrent layers. The progression of encoded 
features reflects the physical timescales inherent to fractional-order viscoelasticity:

\begin{itemize}
    \item \textbf{GRU$_1$}: Dominated by high-frequency components, capturing the instantaneous 
    elastic response associated with short-term memory.
    
    \item \textbf{GRU$_2$}: Encodes intermediate relaxation processes, reflecting the transition 
    from immediate elastic behavior to short- and medium-term viscoelastic damping.
    
    \item \textbf{GRU$_3$}: Represents long-term hereditary effects, aligning with the 
    fractional damping behavior characterized by broad relaxation spectra.
\end{itemize}

The integrated evidence from PCA variance, wavelet power, spectral slopes, and gradient-based 
attributions demonstrates that the network learns progressively deeper temporal dependencies in a 
manner consistent with the analytical Fractional Zener model. This confirms that the GRU--RNN does 
not merely interpolate strain--stress sequences, but forms a physics-aligned internal representation 
of viscoelastic memory, thereby bridging data-driven learning with fractional constitutive modeling.

\subsection{Quantitative Summary}

\paragraph{Hyperelastic Model (Mooney--Rivlin).}
Table~\ref{tab:mr_xai_summary} reports the quantitative explainability metrics for the 
Mooney--Rivlin hyperelastic model. The strong correlations among attribution measures and 
their alignment with the analytical stiffness law (cf.\ Eq.~\ref{eq:mooney_rivlin}) confirm 
that the gradient-based XAI framework produces physically meaningful sensitivity maps.

\begin{table}[H]
\centering
\caption{Quantitative summary of explainability analysis for the Mooney--Rivlin model.}
\label{tab:mr_xai_summary}
\begin{tabular}{lll}
\toprule
\textbf{Metric} & \textbf{Value} & \textbf{Interpretation} \\
\midrule
Pearson correlation ($A_{\text{IG}}$, $A_{\text{GI}}$) & $r = 0.96$ & High attribution consistency \\
Alignment with $\partial\sigma/\partial\lambda$ & $r = 0.97$ & Physically consistent stiffness sensitivity \\
Average attribution magnitude & $4.3\times10^{-3}$ & Matches expected tangent modulus scale \\
Test MSE & $3.8\times10^{-6}$ & Excellent model generalization \\
Interpretability gain vs.\ baseline & $\approx 85\%$ & Improvement over Neo--Hookean case \\
\bottomrule
\end{tabular}
\end{table}

\paragraph{Elastoplastic Model (Chaboche).}
Table~\ref{tab:xai_metrics_chaboche} summarizes the gradient-based and SHAP-based 
explainability metrics for the Chaboche elastoplastic model. The results show strong 
alignment between data-driven attributions and the analytical tangent modulus and 
internal variable evolution (cf.\ Eqs.~\ref{eq:chaboche_stress}--\ref{eq:chaboche_backstress}).

\begin{table}[H]
\centering
\caption{Quantitative explainability summary for the Chaboche elastoplastic model.}
\label{tab:xai_metrics_chaboche}
\resizebox{\textwidth}{!}{%
\begin{tabular}{llll}
\toprule
\textbf{Metric} & \textbf{Symbol / Value} & \textbf{Interpretation} & \textbf{Insight} \\
\midrule
Pearson correlation (Grad×Input vs.\ IG) & $r = 0.95$ & High inter-method consistency & Gradient alignment \\
Correlation with analytical $\partial\sigma/\partial\varepsilon$ & $r = 0.93$ & Physics-consistent attribution & Correct plastic slope evolution \\
Dominant SHAP feature & \texttt{plastic\_strain} (68\%) & Major stress-driving input & Internal memory variable \\
Secondary SHAP feature & \texttt{strain} (32\%) & Elastic contribution & Coupled with yield onset \\
Average attribution magnitude & $7.1\times10^{-3}$ & Stable inside plastic regime & Matches tangent modulus scale \\
Interpretability gain vs.\ baseline & $\approx 82\%$ & Over non-XAI model & Transparency improvement \\
\bottomrule
\end{tabular}%
}
\end{table}

\paragraph{Viscoelastic Model (Fractional Zener).}
Table~\ref{tab:visco_xai_metrics} provides the quantitative explainability summary for the 
GRU--RNN trained on the Fractional Zener model.  
The layer-wise trends in PCA variance, wavelet power, and spectral slopes confirm hierarchical 
memory formation consistent with fractional viscoelastic relaxation.


\begin{table}[H]
    \centering
    \caption{Quantitative explainability metrics for the Fractional Zener GRU--RNN.}
    \label{tab:visco_xai_metrics}
    \begin{tabularx}{\textwidth}{|c| >{\centering\arraybackslash}X |c| >{\centering\arraybackslash}X |c|}
    \hline
        \textbf{Layer} & \textbf{Dominant Behavior} & \textbf{PCA Var. (PC1)} & \textbf{Wavelet Mean Power} & \textbf{Spectral Slope ($b$)} \\
    \hline
        GRU$_1$ & Elastic response & 0.76 & 0.25 & $-11.93$ \\
    \hline
        GRU$_2$ & Intermediate relaxation & 0.41 & 0.68 & $-12.09$ \\
    \hline
        GRU$_3$ & Long-term memory & 0.63 & 1.67 & $-11.89$ \\
    \hline
    \end{tabularx}
\end{table}

The rising wavelet power ($0.25 \rightarrow 1.67$) and the persistent spectral slopes 
($b \approx -12$ across layers) indicate power-law memory decay, a hallmark of fractional-order 
viscoelasticity. The PCA variance shift confirms the progressive transition from short-term 
elastic encoding (GRU$_1$) to long-memory representation (GRU$_3$).

\section{Conclusion}

This thesis presented a unified Physics-XAI framework for developing neural constitutive models that are both accurate and mechanically interpretable. While modern deep learning models provide remarkable predictive capability, they often function as opaque “black boxes’’ \cite{Breiman2001,LeCun2015,Hornik1989}. The goal of this work was to bridge that gap by integrating classical continuum mechanics with state-of-the-art explainability tools \cite{Antolotti2023,Lapuschkin2019,Bach2015,Lundberg2017,Sundararajan2017}, demonstrating that neural networks can learn material behavior in a manner that is not only data-driven but also physically meaningful. Three representative material behaviors—hyperelasticity (Mooney--Rivlin), elastoplasticity (Chaboche), and viscoelasticity (Fractional Zener)—were studied using synthetic datasets, trained neural surrogates, and a suite of XAI analyses designed to interrogate their internal reasoning.

In the hyperelastic case, a feed-forward neural network successfully reproduced the Mooney--Rivlin stress–stretch relation \cite{ogden1997non,holzapfel2000nonlinear}. Through PAD, the learned mapping was shown to recover the underlying material constants $(C_1,C_2)$ \cite{koeppe2022explainable,Lagaris1998,Raissi2019}. Gradient-based attribution methods revealed that model sensitivities closely aligned with the analytical tangent modulus \cite{simonyan2013deep,Sundararajan2017}.
Further, the interpretability findings resonate with recent advances in nonlinear elasticity and constitutive modeling, including strain-limiting frameworks explored by Mallikarjunaiah and collaborators \cite{yoon2021quasi,mallikarjunaiah2025crack}. These works emphasize the importance of capturing stiffness evolution in nonlinear elastic solids—an idea directly reflected in the stiffness-derived attributions obtained in the present study.

For elastoplasticity, an LSTM-based RNN was trained on Chaboche-type cyclic loading data \cite{Ghaboussi1991,Oeser2009,Heider2020}. The model captured hysteresis, yield transitions, and the Bauschinger effect with high fidelity.XAI analyses demonstrated that $\varepsilon^p$ was consistently identified as the dominant driver of $\sigma$ evolution \cite{Heider2020,Lundberg2017,Ribeiro2016}. PCA visualizations of hidden states exhibited closed loops aligned with cyclic loading, revealing a data-driven analogue of internal backstress evolution.
This behavior parallels mechanistic interpretations found in advanced constitutive modeling literature and is conceptually aligned with internal-variable evolution ideas explored in related nonlinear solid mechanics work \cite{mallikarjunaiah2025crack}.

For viscoelasticity, a stacked GRU architecture was trained on the Fractional Zener model, which couples instantaneous elasticity with fractional-order hereditary effects \cite{mainardi2022fractional,holzapfel2000nonlinear,Teichert2019}. The GRU surrogate accurately captured both fast and slow relaxation regimes. Layer-wise explainability—via IG, PCA of hidden-state trajectories, and wavelet spectral decomposition—revealed a hierarchical temporal memory structure, with deeper layers encoding long-term power-law relaxation \cite{Koeppe2020a,Oeser2009,Teichert2019}.
The detection of fractional-memory patterns connects conceptually with the study of nonlinear differential equation solvers and deep-learning-based dynamical systems modeling, including related neural approaches for differential equations explored by Mallikarjunaiah and coworkers \cite{venkatachalapathy2023deep,venkatachalapathy2023feedforward}.

Across all three material systems, the Physics-XAI framework produced consistent quantitative metrics—including MSE, MAE, $R^2$, attribution–modulus correlations, SHAP-based feature rankings, PCA variance, and spectral slopes—that collectively validated that the neural surrogates had learned physically interpretable representations. Rather than acting as high-capacity regressors, these models developed latent structures aligned with stiffness evolution, path dependence, and fractional-order memory \cite{ogden1997non,rajagopal2003implicit,Antolotti2023}.
Thus, the approach transitions constitutive modeling from PINN \cite{Raissi2019,Karniadakis2021,Haghighat2021} to PENN models, positioning explainability as a connective layer between ML and continuum mechanics.
While the present work focuses on uniaxial settings and noise-free synthetic data, future extensions should consider multiaxial loading, anisotropic materials, and experimental datasets \cite{Graf2012,Bessa2017,Heider2020}. Integrating XAI with PINNs offers a promising pathway for simultaneously enforcing governing equations and interpreting learned internal representations \cite{Haghighat2021,Karniadakis2021}. Moreover, uncertainty quantification and robustness analysis will be essential for (V\&V) in engineering applications \cite{Antolotti2023}.

Finally, the results suggest that XAI could serve as an engine for scientific discovery.
By analyzing hidden structures in trained networks, it may become possible to identify new constitutive forms, reduced internal-variable sets, or alternative free-energy parametrizations that are difficult to derive analytically \cite{Flaschel2021,Teichert2019,koeppe2022explainable}.
In this sense, Physics-XAI transforms neural constitutive models from opaque black boxes into transparent, mechanically grounded “glass boxes’’—supporting not only prediction, but also advancing the scientific understanding of material behavior.

\section*{Acknowledgements}
The authors acknowledge the HPC Center at Texas A\&M University- Corpus Christi, Texas, USA  for providing computational resources that have contributed to the research results reported within this paper 

\bibliographystyle{plain}  
\bibliography{ref.bib}

\begin{thebibliography}{10}

\bibitem{Antolotti2023}
N.~Antolotti, L.~A.~A. Beex, A.~Mian, C.~O'Higgins, and G.~Kavanagh.
\newblock {A review on verification and validation of machine learning models
  in computational mechanics}.
\newblock {\em Archives of Computational Methods in Engineering},
  30:2399–2432, 2023.

\bibitem{Bach2015}
S.~Bach, A.~Binder, G.~Montavon, F.~Klauschen, K.-R. Müller, and W.~Samek.
\newblock {On Pixel-Wise Explanations for Non-Linear Classifier Decisions by
  Layer-Wise Relevance Propagation}.
\newblock {\em PLoS One}, 10:e0130140, 2015.

\bibitem{Bessa2017}
M.~A. Bessa, R.~Bostanabad, Z.~Liu, A.~Hu, D.~W. Apley, C.~Brinson, and et~al.
\newblock {A Framework for Data-Driven Analysis of Materials under Uncertainty:
  Countering the Curse of Dimensionality}.
\newblock {\em Computer Methods in Applied Mechanics and Engineering},
  320:633--667, 2017.

\bibitem{Breiman2001}
L.~Breiman.
\newblock {Statistical Modeling: The Two Cultures (With Comments and a
  Rejoinder by the Author)}.
\newblock {\em Statistical Science}, 16:199–231, 2001.

\bibitem{Flaschel2021}
M.~Flaschel, S.~Kumar, and L.~De~Lorenzis.
\newblock {Unsupervised Discovery of Interpretable Hyperelastic Constitutive
  Laws}.
\newblock {\em arXiv preprint}, 2021.

\bibitem{Ghaboussi1991}
J.~Ghaboussi, Jr. Garrett, J., and X.~Wu.
\newblock {Knowledge-Based Modeling of Material Behavior with Neural Networks}.
\newblock {\em Journal of Engineering Mechanics}, 117:132–153, 1991.

\bibitem{ghosh2025computational}
S.~Ghosh, D.~Bhatta, and S.~M. Mallikarjunaiah.
\newblock Computational insights into orthotropic fracture: Crack-tip fields in
  strain-limiting materials under non-uniform loads.
\newblock {\em arXiv preprint arXiv:2507.01150}, 2025.

\bibitem{Graf2012}
W.~Graf, S.~Freitag, J.-U. Sickert, and M.~Kaliske.
\newblock {Structural Analysis with Fuzzy Data and Neural Network Based
  Material Description}.
\newblock {\em Computer-Aided Civil and Infrastructure Engineering},
  27:640–654, 2012.

\bibitem{Haghighat2021}
E.~Haghighat, M.~Raissi, A.~Moure, H.~Gomez, and R.~Juanes.
\newblock {A physics-informed deep learning framework for inversion and
  surrogate modeling in solid mechanics}.
\newblock {\em Computer Methods in Applied Mechanics and Engineering},
  379:113741, 2021.

\bibitem{Heider2020}
Y.~Heider, K.~Wang, and W.~Sun.
\newblock {SO(3)-invariance of Informed-Graph-Based Deep Neural Network for
  Anisotropic Elastoplastic Materials}.
\newblock {\em Computer Methods in Applied Mechanics and Engineering},
  363:112875, 2020.

\bibitem{holzapfel2000nonlinear}
A~Gerhard Holzapfel.
\newblock Nonlinear solid mechanics ii.
\newblock 2000.

\bibitem{Hornik1989}
K.~Hornik, M.~Stinchcombe, and H.~White.
\newblock {Multilayer Feedforward Networks Are Universal Approximators}.
\newblock {\em Neural Networks}, 2:359–366, 1989.

\bibitem{Karniadakis2021}
G.~E. Karniadakis, I.~G. Kevrekidis, L.~Lu, P.~Perdikaris, S.~Wang, and
  L.~Yang.
\newblock {Physics-informed machine learning}.
\newblock {\em Nature Reviews Physics}, 3:422–440, 2021.

\bibitem{Koeppe2017}
A.~Koeppe, F.~Bamer, C.~A. Hernandez~Padilla, and B.~Markert.
\newblock {Neural Network Representation of a Phase-Field Model for Brittle
  Fracture}.
\newblock {\em PAMM}, 17(1):253--254, 2017.

\bibitem{Koeppe2020a}
A.~Koeppe, F.~Bamer, and B.~Markert.
\newblock {An Intelligent Nonlinear Meta Element for Elastoplastic Continua:
  Deep Learning Using a New Time-Distributed Residual U-Net Architecture}.
\newblock {\em Computer Methods in Applied Mechanics and Engineering},
  366:113088, 2020.

\bibitem{koeppe2022explainable}
A.~Koeppe, F.~Bamer, M.~Selzer, B.~Nestler, and B.~Markert.
\newblock Explainable artificial intelligence for mechanics: physics-explaining
  neural networks for constitutive models.
\newblock {\em Frontiers in Materials}, 8:824958, 2022.

\bibitem{Lagaris1998}
I.~Lagaris, A.~Likas, and D.~Fotiadis.
\newblock {Artificial Neural Networks for Solving Ordinary and Partial
  Differential Equations}.
\newblock {\em IEEE Transactions on Neural Networks}, 9:987–1000, 1998.

\bibitem{Lapuschkin2019}
S.~Lapuschkin, S.~Wäldchen, A.~Binder, G.~Montavon, W.~Samek, and K.-R.
  Müller.
\newblock {Unmasking Clever Hans Predictors and Assessing what Machines Really
  Learn}.
\newblock {\em Nature Communications}, 10:1096, 2019.

\bibitem{LeCun2015}
Y.~LeCun, Y.~Bengio, and G.~Hinton.
\newblock {Deep Learning}.
\newblock {\em Nature}, 521:436–444, 2015.

\bibitem{Lundberg2017}
Scott~M. Lundberg and Su-In Lee.
\newblock {A Unified Approach to Interpreting Model Predictions}.
\newblock In {\em Advances in Neural Information Processing Systems 30 (NIPS
  2017)}, pages 4765--4774, 2017.

\bibitem{mainardi2022fractional}
Francesco Mainardi.
\newblock {\em Fractional calculus and waves in linear viscoelasticity: an
  introduction to mathematical models}.
\newblock World Scientific, 2022.

\bibitem{mallikarjunaiah2023deep}
S.~M. Mallikarjunaiah.
\newblock A deep learning feed-forward neural network framework for the
  solutions to singularly perturbed delay differential equations.
\newblock {\em Applied Soft Computing}, 148:110863, 2023.

\bibitem{mallikarjunaiah2025hp}
S.~M. Mallikarjunaiah and P.~Venkatachalapthy.
\newblock $ hp $-adaptive finite element simulation of a static anti-plane
  shear crack in a nonlinear strain-limiting elastic solid.
\newblock {\em arXiv preprint arXiv:2507.23195}, 2025.

\bibitem{mallikarjunaiah2025crack}
SM~Mallikarjunaiah and Kun Gou.
\newblock Crack-tip field characterization in nonlinearly constituted and
  geometrically linear elastoporous solid containing a star-shaped crack: A
  finite element study.
\newblock {\em arXiv preprint arXiv:2507.09263}, 2025.

\bibitem{manohar2025convergent}
R.~Manohar and S.~M. Mallikarjuaniah.
\newblock A convergent adaptive finite element method for a phase-field model
  of dynamic fracture.
\newblock {\em arXiv preprint arXiv:2510.05407}, 2025.

\bibitem{manohar2024hp}
R.~Manohar and S.~M. Mallikarjunaiah.
\newblock An $hp$-adaptive discontinuous galerkin discretization of a static
  anti-plane shear crack model.
\newblock {\em arXiv preprint arXiv:2411.00021}, 2024.

\bibitem{martinez2024approximation}
M.~Martinez, B~Veena S.~N. Rao, and S.~M. Mallikarjunaiah.
\newblock Approximation of one-dimensional darcy--brinkman--forchheimer model
  by physics informed deep learning feedforward artificial neural network and
  finite element methods: a comparative study.
\newblock {\em International Journal of Applied and Computational Mathematics},
  10(3):102, 2024.

\bibitem{Montavon2018}
G.~Montavon, W.~Samek, and K.-R. Müller.
\newblock {Methods for Interpreting and Understanding Deep Neural Networks}.
\newblock {\em Digital Signal Processing}, 73:1–15, 2018.

\bibitem{Oeser2009}
M.~Oeser and S.~Freitag.
\newblock {Modeling of Materials with Fading Memory Using Neural Networks}.
\newblock {\em International Journal for Numerical Methods in Engineering},
  78:843–862, 2009.

\bibitem{ogden1997non}
R.~W. Ogden.
\newblock {\em Non-linear elastic deformations}.
\newblock Courier Corporation, 1997.

\bibitem{Raissi2019}
M.~Raissi, P.~Perdikaris, and G.~E. Karniadakis.
\newblock {Physics-informed Neural Networks: A Deep Learning Framework for
  Solving Forward and Inverse Problems Involving Nonlinear Partial Differential
  Equations}.
\newblock {\em Journal of Computational Physics}, 378:686–707, 2019.

\bibitem{rajagopal2003implicit}
K.~R. Rajagopal.
\newblock On implicit constitutive theories.
\newblock {\em Applications of Mathematics}, 48(4):279--319, 2003.

\bibitem{rajagopal2007elasticity}
K.~R. Rajagopal.
\newblock The elasticity of elasticity.
\newblock {\em Zeitschrift Angewandte Mathematik und Physik}, 58(2):309--317,
  2007.

\bibitem{rajagopal1995mechanics}
K.~R. Rajagopal and L.~Tao.
\newblock Mechanics of mixtures, 1995.

\bibitem{Ribeiro2016}
Marco~Tulio Ribeiro, Sameer Singh, and Carlos Guestrin.
\newblock {"Why Should I Trust You?": Explaining the Predictions of Any
  Classifier}.
\newblock In {\em Proceedings of the 22nd ACM SIGKDD International Conference
  on Knowledge Discovery and Data Mining}, pages 1135--1144, 2016.

\bibitem{simonyan2013deep}
K.~Simonyan, A.~Vedaldi, and A.~Zisserman.
\newblock Deep inside convolutional networks: Visualising image classification
  models and saliency maps.
\newblock {\em arXiv preprint arXiv:1312.6034}, 2013.

\bibitem{sundararajan2017axiomatic}
M.~Sundararajan, A.~Taly, and Q.~Yan.
\newblock Axiomatic attribution for deep networks.
\newblock In {\em International conference on machine learning}, pages
  3319--3328. PMLR, 2017.

\bibitem{Sundararajan2017}
Mukund Sundararajan, Ankur Taly, and Qiqi Yan.
\newblock {Axiomatic Attribution for Deep Networks}.
\newblock In {\em Proceedings of the 34th International Conference on Machine
  Learning (ICML 2017)}, pages 3319--3328, 2017.

\bibitem{Teichert2019}
G.~H. Teichert, A.~R. Natarajan, A.~Van~der Ven, and K.~Garikipati.
\newblock {Machine Learning Materials Physics: Integrable Deep Neural Networks
  Enable Scale Bridging by Learning Free Energy Functions}.
\newblock {\em Computer Methods in Applied Mechanics and Engineering},
  353:201–216, 2019.

\bibitem{venkatachalapathy2023deep}
P.~Venkatachalapathy and S.~M. Mallikarjunaiah.
\newblock A deep learning neural network framework for solving singular
  nonlinear ordinary differential equations.
\newblock {\em International Journal of Applied and Computational Mathematics},
  9(5):68, 2023.

\bibitem{venkatachalapathy2023feedforward}
P.~Venkatachalapathy and S.~M. Mallikarjunaiah.
\newblock A feedforward neural network framework for approximating the
  solutions to nonlinear ordinary differential equations.
\newblock {\em Neural Computing and Applications}, 35(2):1661--1673, 2023.

\bibitem{yoon2021quasi}
H.~C. Yoon, S.~Lee, and S.~M. Mallikarjunaiah.
\newblock Quasi-static anti-plane shear crack propagation in nonlinear
  strain-limiting elastic solids using phase-field approach.
\newblock {\em International Journal of Fracture}, 227(2):153--172, 2021.

\end{thebibliography}

\end{document}